\documentclass[12pt,british,english,onecolumn]{IEEEtran}
\usepackage[T1]{fontenc}
\usepackage[latin9]{inputenc}
\usepackage{geometry}
\usepackage{babel}
\usepackage{verbatim}
\usepackage{float}
\usepackage{textcomp}
\usepackage{amsthm}
\usepackage{amsmath}
\usepackage{amssymb}
\usepackage{graphicx}
\usepackage{setspace}
\usepackage{esint}
\usepackage[unicode=true]
 {hyperref}
\usepackage{breakurl}

\makeatletter

\newcommand{\lyxmathsym}[1]{\ifmmode\begingroup\def\b@ld{bold}
  \text{\ifx\math@version\b@ld\bfseries\fi#1}\endgroup\else#1\fi}

\theoremstyle{plain}
\newtheorem{thm}{\protect\theoremname}
\theoremstyle{definition}
\newtheorem{defn}[thm]{\protect\definitionname}
\theoremstyle{definition}
\newtheorem{example}[thm]{\protect\examplename}

\usepackage{fancyhdr}
\usepackage{lettrine}
\usepackage{caption2}
\usepackage{tocloft}   

\@ifundefined{showcaptionsetup}{}{%
 \PassOptionsToPackage{caption=false}{subfig}}
\usepackage{subfig}
\makeatother

\addto\captionsbritish{\renewcommand{\definitionname}{Definition}}
\addto\captionsbritish{\renewcommand{\examplename}{Example}}
\addto\captionsbritish{\renewcommand{\theoremname}{Theorem}}
\addto\captionsenglish{\renewcommand{\definitionname}{Definition}}
\addto\captionsenglish{\renewcommand{\examplename}{Example}}
\addto\captionsenglish{\renewcommand{\theoremname}{Theorem}}
\providecommand{\definitionname}{Definition}
\providecommand{\examplename}{Example}
\providecommand{\theoremname}{Theorem}

\begin{document}

\title{\noindent On Pseudocodewords and Improved Union Bound of Linear Programming
Decoding of HDPC Codes}

\author{\IEEEauthorblockN{Ohad~Gidon and Yair~Be'ery~\emph{Senior}~\emph{Member},~\emph{IEEE}}
\IEEEauthorblockA{\\
 Tel Aviv University, School of Electrical Engineering \\
 Ramat Aviv 69978, ISRAEL\\
 Email: \href{http://ohadgodo@post.tau.ac.il}{ohadgodo@post.tau.ac.il},
\href{http://ybeery@eng.tau.ac.il}{ybeery@eng.tau.ac.il}}}

\maketitle

\begin{abstract}
\fontsize{10}{11}\selectfont \textnormal  In this paper, we present
an improved union bound on the Linear Programming (LP) decoding performance
of the binary linear codes transmitted over an additive white Gaussian
noise channels. The bounding technique is based on the second-order
of Bonferroni-type inequality in probability theory, and it is minimized
by Prim's minimum spanning tree algorithm. The bound calculation needs
the fundamental cone generators of a given parity-check matrix rather
than only their weight spectrum, but involves relatively low computational
complexity. It is targeted to high-density parity-check codes, where
the number of their generators is extremely large and these generators
are spread densely in the Euclidean space. We explore the generator
density and make a comparison between different parity-check matrix
representations. That density effects on the improvement of the proposed
bound over the conventional LP union bound. The paper also presents
a complete pseudo-weight distribution of the fundamental cone generators
for the BCH{[}31,21,5{]} code.\end{abstract}

\section{Introduction\label{Introduction} }

\IEEEPARstart{T}{he} calculation of error probability for Linear
Programming (LP) decoding of Binary Phase-Shift Keying (BPSK) modulated
binary codes is often a complex task. This is mainly due to the complexity
of LP Voronoi or decision regions \cite{DecisionRegions} \cite{GraphCover}.
The probability of correct decision in an Additive White Gaussian
Noise (AWGN) channel, can be obtained by integrating a multidimensional
Gaussian distribution over the decision region of the transmitted
codeword (CW). 

LP decoding is a relaxed version of the Maximum-Likelihood (ML) decoding.
The \emph{codeword polytope} \cite{UsingLP} of ML is replaced by
a relaxed polytope, called the \emph{fundamental polytope} \cite{UsingLP}.
The fundamental polytope arisen from a given parity check matrix.
Its vertices are every codeword, but it also has some non-codeword.
The vertices of the codeword polytope are the all codewords, and the
vertices of the fundamental polytope are called \emph{pseudocodewords
}(PCWs) \cite{UsingLP}. The additional non-codewords make the decision
region \cite{DecisionRegions} of the LP decoder even more complex
than that of the ML. Therefore, a derivation of analytical bounds
has an important role in evaluating the performance of the LP decoder.

The \emph{fundamental cone} \cite{GraphCover} is the conic hull of
the fundamental polytope. The LP error probability over the fundamental
polytope is equal to that over the fundamental cone \cite{RelaxationBounds}.
Moreover, it is sufficient to consider\emph{ }only the fundamental
cone\emph{ generators }\cite{RelaxationBounds} for evaluating the
performance of the LP decoder.

The well-known upper bound on the error probability of a digital communication
system is the \emph{Union Bound} (UB), which is a first-order \emph{Bonferroni-type}
inequality \cite{Bonferroni} in the probability theory. The UB of
the LP decoder \cite{DecisionRegions} \cite{LP-UB} \cite{LP-UB2}
for High-Density Parity-Check (HDPC) codes presets inaccurate results
due to the high density of fundamental cone generators. In fact, the
union bound sums all of the pairwise error events as if they were
disjoint, but this scenario is far from being the case in LP decoding
of HDPC codes.

Each pseudocodeword in the LP decoder can be located in the BPSK signal
space \cite{GraphCover}. What the LP decoder does, it chooses the
nearest pseudocodeword to the received vector as the most likely transmitted
pseudocodeword. The ML soft decision decoder has such property as
well, but unlike to the LP decoder, its signal space contains only
the set of the all codewords. Thus many of ML upper bounds can be
reused \cite{TSB} \cite{SphereBound} \cite{NewTSB} \cite{ITSB}
in the case of LP decoding.

For a given code, each of its parity-check matrix creates a fundamental
cone with different pseudo-weight spectrum and geometrical structure,
which influences differently on the error probability of the LP decoder.
Therefore, the geometrical properties of the fundamental cone generators
are essential to evaluate with a better accuracy the LP decoding error
probability. Thus ML error probability bounds which use the weight
spectrum of the code or those who sum the error contribution of each
individual codeword become less attractive. In \cite{ITSB} a ML bound
is presented which is based on the second-order upper bound on the
probability of a finite union of events. And indeed, it uses the geometrical
properties of the codewords and considers an intersection of pairwise
error events, but involves relatively high computational complexity.

To explore the density of the fundamental cone generators, we have
defined the \emph{angle graph}: each generator is considered as a
node of a complete undirected graph. The cost of an edge is the angle
between the generators related to the adjacent nodes. The minimum
spanning tree is found and its cost distribution is illustrated. Different
patterns for various parity-check matrices were observed.

In this paper, we propose an upper bound based on the second-order
of Bonferroni-type inequality. The bound needs the fundamental cone
generators rather than their weight spectrum. We call it Improved
Linear Programming Union Bound (ILP-UB). It consists of two parts:
The first term is the LP union bound itself, and the second term is
a second-order correction that can be optimized by a known minimum
spanning tree algorithm. It requires relatively low computational
complexity since it involves only the $Q$-function.

The proposed ILP-UB makes use of an upper bound of the triplet-wise
error probability that has been introduced earlier in the paper. We
derive analytical expression to evaluate the triple-wise error probability
depending on the angle which they create. And for example, the triple-wise
error probability for the minimal-weight generators of the BCH{[}63,57,3{]}
code is calculated. It is compared to the triple-wise error upper
bound and to the UB in different angles and Signal-to-Noise Ratios
(SNRs).

The proposed ILP-UB was tested on three HDPC codes: Golay{[}24,12,8{]},
BCH{[}31,26,3{]}, BCH{[}63 ,57,3{]}, and on the Low-Density Parity-Check
(LDPC) Tanner code {[}155,64,20{]} \cite{Tanner155}. An improvement
of up to 0.37 dB has been demonstrated over the conventional Linear
Programming Union Bound (LP-UB).

This paper is organized as follows. Sec. \ref{Preliminaries} provides
some background on ML and LP decoding. The minimum spanning tree problem
for undirected graph is also reviewed in Sec. \ref{Preliminaries}.
In Sec. \ref{Density} we explore the density of the fundamental cone
generators and we check the effect of that density on the union bound
of the triplet-wise error probability. The problem of finding an LP
dominant error events is discussed in Sec. \ref{DominantGroup}. In
Sec. \ref{ImprovedUB} we propose an improved linear programming error
union bound. Sec. \ref{Results} provides numerical results and discusses
some possible direction for further research on how to improve the
proposed bound. Sec. \ref{Conclusions} concludes the paper. 

\renewcommand{\figurename}{\fontsize{9}{10}\selectfont  Fig.}

\renewcommand\captionlabeldelim{.  }

\section{Preliminaries and Definitions\label{Preliminaries} }

\subsection{ML and LP Decoding}

In this section we briefly review ML and LP decoding \cite{UsingLP}.
Consider a binary linear code $\mathcal{C}$ of length $n$, dimension
$k$ and code rate $R\triangleq k/n$. Let $\mathbb{F}_{2}\triangleq\{0,1\}$
denote the finite field with two elements. The code $\mathcal{C}$
is defined by some $m\times n$ parity-check matrix $H\in\mathbb{F}_{2}^{m\mathrm{x}n}$
with row vectors $\mathrm{\mathbf{h}}_{1},\mathrm{\mathbf{h}}_{2},...,\mathrm{\mathbf{h}}_{m}$,
i.e. $\mathcal{C\triangleq}\{\mathrm{\mathbf{x}}\in\mathbb{F}_{2}^{n}\mid\mathrm{\mathbf{x}}H^{T}=0\}$.
The code will be called an {[}\emph{n,k,d}{]} code, in which \emph{d}
is its minimum \emph{Hamming distance}. The code is used for data
communication over a memoryless binary-input channel with channel
law $P_{Y|X}(y|x)$. We denote the transmitted codeword by $\mathrm{\mathbf{x}}\triangleq(x_{1},...,x_{n})$,
the transmitted signal by $\overline{\mathrm{\mathbf{x}}}\triangleq(\overline{x}_{1},...,\overline{x}_{n})$
and the received signal by $\mathrm{\mathbf{y}}\triangleq(y_{1},...,y_{n})$.
We assume that every codeword $\mathrm{\mathbf{x}}\in\mathcal{C}$
is transmitted with equal probability. Let $\boldsymbol{\lambda}$
denote the Log-Likelihood Ratio (LLR) vector with the LLR components
$\lambda_{i}\triangleq P_{Y|X}(y_{i}|0)/P_{Y|X}(y_{i}|1)$ for $i=1,...,n$.
The block-wise Maximum Likelihood Decoding (MLD) is \vspace{-7mm}

\begin{equation}
\hat{\mathrm{\mathbf{x}}}_{MLD}(\mathrm{\mathbf{y}})\triangleq\underset{\mathrm{\mathbf{x}}\in\mathcal{C}}{\mathrm{arg\: min}}\left\langle \mathrm{\mathbf{x}},\boldsymbol{\lambda}\right\rangle .\label{eq:MLD}
\end{equation}
Where $\left\langle \mathrm{\mathbf{x}},\boldsymbol{\lambda}\right\rangle \triangleq\sum_{i}x_{i}\lambda_{i}$
denote the\emph{ }standard\emph{ inner} \emph{product} of two vectors
of equal length. The ML decoder error probability is independent of
the transmitted CW, therefore, we assume without loss of generality
that the all-zeros codeword $\mathrm{\mathbf{x}}_{0}$ is transmitted.
Then \cite{CommunBook} \vspace{-8mm}

\begin{eqnarray}
P_{r}^{MLD}(error\mid\mathrm{\mathbf{x}}_{0}) & = & P_{r}\left(\hat{\mathrm{\mathbf{x}}}_{MLD}(\mathrm{\mathbf{y}})\neq\mathrm{\mathbf{x}}_{0}\mid\mathrm{\mathbf{x}}_{0}\right)\\
 & = & P_{r}\left\{ \bigcup_{\mathrm{x}\in\mathcal{C}\setminus\mathrm{\mathbf{x}}_{0}}||\overline{\mathrm{\mathbf{x}}}-\mathrm{\mathbf{y}}||_{2}\leq|||\mathrm{\overline{\mathbf{x}}}_{0}-\mathrm{\mathrm{\mathbf{y}}}||_{2}\mid\mathrm{\mathbf{x}}_{0}\right\} \label{eq:ML error}\\
 & \leq & \sum_{\mathrm{x}\in\mathcal{C}\setminus\mathrm{x}_{0}}P_{r}\left\{ \,||\overline{\mathrm{\mathbf{x}}}-\mathrm{\mathrm{\mathbf{y}}}||_{2}\leq|||\mathrm{\overline{\mathbf{x}}}_{0}-\mathrm{\mathrm{\mathbf{y}}}||_{2}\mid\mathrm{\mathbf{x}}_{0}\,\right\} \\
 & = & \sum_{\mathrm{x}\in\mathcal{C}\setminus\mathrm{\mathbf{x}}_{0}}Q\left(\dfrac{d_{\mathrm{\mathbf{x}}}}{2\sigma}\right).\label{eq:ML-UB}
\end{eqnarray}
Where the $Q$-function is defined to be $Q(x)\triangleq\frac{1}{\sqrt{2\pi}}\int_{x}^{\infty}\exp\Bigl(-\frac{t^{2}}{2}\Bigr)dt$
and $||\mathrm{\mathbf{x}}||_{2}\triangleq\sqrt{\sum_{i}x_{i}^{2}}$
denote the $\mathcal{L}_{2}$-norm of a vector \textbf{x}. Eq. \eqref{eq:ML error}
also allows to make a simulation of the error probability contributed
by a subgroup of codewords. Eq. \eqref{eq:ML-UB} is the ML union
bound, where $d_{\mathrm{\mathbf{x}}}\triangleq||\overline{\mathrm{\mathbf{x}}}-\mathrm{\overline{\mathbf{x}}}_{0}||_{2}=2\sqrt{RE_{b}w_{H}(\mathrm{\mathbf{x}})}$
is the Euclidean distance from $\overline{\mathrm{\mathbf{x}}}$ to
the transmitted signal $\mathrm{\overline{\mathbf{x}}}_{0}$. 

The MLD \eqref{eq:MLD} can be formulated \cite{UsingLP} as the following
equivalent optimization problem:\vspace{-6mm}

\begin{equation}
\hat{\mathrm{\mathbf{x}}}_{MLD}(\mathrm{\mathbf{y}})\triangleq\underset{\mathrm{\mathbf{x}}\in conv(\mathcal{C})}{\mathrm{arg}\:\mathrm{min}}\left\langle \mathrm{\mathbf{x}},\boldsymbol{\lambda}\right\rangle .\label{eq:OriginalLP}
\end{equation}
$conv(\mathcal{C})$ is called the \emph{codeword polytope} \cite{UsingLP},
which is the convex hull of all possible codewords. The vertices of
the codeword polytope are the all codewords. The number of inequalities
needed to describe it grows exponentially in the code length. Therefore,
solving this linear programming problem is not practical for codes
with reasonable block length. To make this problem more feasible it
was suggested \cite{UsingLP} to replace $conv(\mathcal{C})$ by a
relaxed polytope $\mathcal{P}\triangleq\mathcal{P}(H)$, called the
\emph{fundamental polytope. }\vspace{-4mm}\emph{ }
\begin{eqnarray*}
\mathrm{\mathcal{P}\triangleq\overset{m}{\underset{\mathit{j}=1}{\bigcap}}\, conv}(\mathcal{C}_{\text{\ensuremath{j}}})\qquad\mathrm{with}\quad\mathcal{C}_{\text{\ensuremath{j}}} & \triangleq & \left\{ \mathrm{\mathbf{x}}\in\mathbb{F}_{2}^{n}\mid\mathrm{\mathbf{x}}\mathrm{\mathbf{h}}_{j}^{T}=0\right\} .
\end{eqnarray*}
Where $conv(\mathcal{C})\subseteq conv(\mathcal{C}_{\text{\ensuremath{j}}})$
for $\text{\ensuremath{j}}=1,...,m$ and hence $conv(\mathcal{C})\subseteq\mathcal{P}(H)\subset[0,1]^{n}$.
The number of inequalities that describe $\mathcal{P}(H)$ is typically
much smaller than those of $conv(\mathcal{C})$. The Linear Programming
Decoding (LPD) is then \vspace{-6mm}

\begin{equation}
\hat{\boldsymbol{\omega}}_{LPD}(\mathrm{\mathbf{y}})\triangleq\underset{\boldsymbol{\omega}\in\mathcal{P}}{\mathrm{arg\:}\mathrm{min}}\left\langle \boldsymbol{\omega},\boldsymbol{\lambda}\right\rangle .
\end{equation}
 In the case of $conv(\mathcal{C})=\mathcal{P}(H)$ the relaxed LP
solution equals to that of ML. In the case of $conv(\mathcal{C})\subset\mathcal{P}(H)$
the relaxed LP problem represents a suboptimal decoder which has vertices
in $\mathcal{P}(H)$ which are not in $conv(\mathcal{C})$. The vertices
of $\mathcal{P}(H)$, denoted by $\mathcal{V}(\mathcal{P}(H))$, are
called LP pseudocodewords. 

The \emph{fundamental cone} \cite{GraphCover} $\mathcal{K}(H)\triangleq\mathcal{K}$
is defined to be the conic hull of the fundamental polytope i.e. the
set that consists of all possible conic combinations of all the points
in $\mathcal{P}(H)$ and hence $\mathcal{P}(H)\subset\mathcal{K}(H)$.
The LP decoding error probability over the fundamental polytope is
equal to that over the fundamental cone \cite{RelaxationBounds}.
We let $\mathcal{\mathbb{R}}$ and $\mathbb{R_{\mathrm{+}}}$ be the
set of real numbers and the set of non-negative real numbers, respectively.

\medskip{}

\begin{defn}
\label{def:generators}( \cite{RelaxationBounds}, \cite{LiftingFund})
A set $\mathcal{G}(\mathcal{\mathcal{K}})\triangleq\{\mathrm{\mathbf{g}}_{1},\mathrm{\mathbf{g}}_{2},...,\mathrm{\mathbf{g}}_{M}\mid\mathbf{g}_{i}\in\mathbb{R}_{+}^{n},\; i=1,...,M\}$
of \emph{M} linearly independent vectors where $\mathcal{\mathcal{K}}=\left\{ \overset{M}{\underset{i=1}{\sum}}\alpha_{i}\mathrm{\mathbf{g}}_{i}\mid\alpha_{i}\in\mathbb{R}\right\} $
are called the \emph{generators} of the cone $\mathcal{\mathcal{K}}$.
\qed 
\end{defn}
\bigskip{}
It follows from Def. \ref{def:generators} that a vector $\mathrm{\mathbf{x}}$
is in $\mathcal{K}$ if and only if $\mathrm{\mathbf{x}}$ can be
written as a nonnegative linear combination of the generators, i.e.
$\mathrm{\mathbf{x}}=\overset{M}{\underset{i=1}{\sum}}\alpha_{i}\mathrm{\mathrm{\mathbf{g}}}_{i}$
where $\alpha_{i}\in\mathbb{R}$. Note that a set of generators is
not unique, and that the all-zeros codeword $\mathrm{\mathrm{\mathbf{x}}}_{0}\notin\mathcal{G}(\mathcal{\mathcal{K}}).$

We assume an AWGN channel, where each \emph{i}-th transmitted bit
perturbed by a white Gaussian noise $z_{i}$ with a zero mean and
noise power $\sigma^{2}\triangleq N_{0}/2$. The received signal is
$\mathbf{\mathrm{\mathbf{y}}}=\overline{\mathrm{\mathbf{x}}}+\mathrm{\mathbf{z}}$,
where $\mathrm{\mathbf{z}}$ designates an\emph{ n}-dimensional Gaussian
noise vector with independent components $z_{1},z_{2},...,z_{n}$. 

We consider a BPSK modulation: the transmitted signal is $\overline{\mathrm{\mathbf{x}}}=\gamma\left(1-2\mathrm{\mathbf{x}}\right)$,
where $\gamma\triangleq\sqrt{RE_{b}}$ in which $E_{b}$ is the information
bit energy. The signal-to-noise ratio is defined to be $\mathrm{SNR}\triangleq E_{b}/N_{0}$.
Following from the above, the LLR vector is $\boldsymbol{\lambda}=4\frac{\sqrt{RE_{b}}}{N_{0}}\mathrm{\mathbf{y}}$
\cite{GraphCover}, and therefore, the LPD will be considered henceforth
\vspace{-10mm}

\begin{equation}
\hat{\boldsymbol{\omega}}_{LPD}=\underset{\boldsymbol{\omega}\in\mathcal{P}}{\mathrm{arg\:}\mathrm{min}}\left\langle \boldsymbol{\omega},\mathrm{\mathbf{y}}\right\rangle .
\end{equation}

\begin{defn}
(\cite{GraphCover}, \cite{PhDWiberg}, \cite{EffectiveWeights})
Let $\boldsymbol{\omega}\in\mathbb{R}_{+}^{n}.$ The AWGN channel
pseudo-weight $w_{p}^{AWGNC}(\boldsymbol{\omega})$ of $\boldsymbol{\omega}$
is given by \vspace{-10mm}
\end{defn}
\begin{equation}
w_{p}^{AWGNC}(\boldsymbol{\omega})\triangleq\frac{||\boldsymbol{\omega}||_{1}^{2}}{||\boldsymbol{\omega}||_{2}^{2}}\text{,}
\end{equation}
where $||\mathrm{\mathbf{x}}||_{1}\triangleq\sum_{i}|x_{i}|$ denote
the $\mathcal{L}_{1}$-norm of a vector x. If $\boldsymbol{\omega}=0$
we define $w_{p}^{AWGNC}(\boldsymbol{\omega})\triangleq0$, and in
the case of $\boldsymbol{\omega}\in\{0,1\}^{n}$ we have $w_{p}^{AWGNC}(\boldsymbol{\omega})$
= $w_{H}(\boldsymbol{\omega})$\emph{.}\qed 

\bigskip{}

\noindent For an easier notation, as we discuss in this paper only
AWGN channel, we will use the shorter notation $w_{p}(\boldsymbol{\omega})$
instead of $w_{p}^{AWGNC}(\boldsymbol{\omega})$.

\medskip{}

Due to the symmetry property of the fundamental polytope the probability
that the LP decoder fails is independent of the codeword that was
transmitted \cite{UsingLP}. Therefore, we henceforth assume without
loss of generality when analyzing LPD error probability, that the
all-zeros codeword $\mathrm{\mathbf{x}}_{0}$ is transmitted. 

The set of optimal solutions of a closed convex LP problem always
includes at least one vertex of the polytope. Therefore, the LPD error
probability is \vspace{-5mm}

\begin{equation}
P_{r}^{LPD}(error\mid\mathrm{\mathbf{x}}_{0})=P_{r}\left\{ \bigcup_{\boldsymbol{\omega}\in\mathcal{V}(\mathcal{\mathcal{P}}(H))\setminus\mathrm{\mathbf{x}}_{0}}\left\langle \boldsymbol{\omega},\mathrm{\mathbf{y}}\right\rangle \leq0\mid\mathrm{\mathbf{x}}_{0}\right\} .\label{eq:LP-errorP}
\end{equation}
A pseudocodeword $\mathrm{\mathbf{p}}\in\mathcal{V}(\mathcal{P})$
also belongs to the fundamental cone. Thus it can be written as a
non-negative linear combination of the generators, i.e. $\mathrm{\mathbf{p}}=\overset{M}{\underset{i=1}{\sum}}\alpha_{i}\mathrm{\mathrm{\mathbf{g}}}_{i}$
with $\alpha_{i}\geq0$. Therefore, if there is $\mathrm{\mathbf{p}}\in\mathcal{V}(\mathcal{P})$
such that $\left\langle \mathrm{\mathbf{p}},\mathrm{\mathbf{y}}\right\rangle =\overset{M}{\underset{i=1}{\sum}}\alpha_{i}\left\langle \mathrm{\mathrm{\mathbf{g}}}_{i},\mathrm{\mathbf{y}}\right\rangle <0$,
then there must be at least one generator $\mathrm{\mathbf{g}}_{i}\mathcal{\in G}(\mathcal{\mathcal{K}})$
such that $\left\langle \mathrm{\mathrm{\mathbf{g}}}_{i},\mathrm{\mathbf{y}}\right\rangle <0.$
Therefore, the union of the pseudocodewords' error events in \eqref{eq:LP-errorP}
can be replaced by the union of the generators' error events.

A vector $\boldsymbol{\omega}\in\mathbb{R}_{+}^{n}$ which is not
codeword can be located into the signal space in the same way as a
codeword, i.e $\mathrm{\boldsymbol{\overline{\omega}}}=\gamma\left(1-2\mathrm{\mathbf{\boldsymbol{\omega}}}\right)$.
The vector $\mathrm{\boldsymbol{\omega}}_{virt}\triangleq\tfrac{||\mathrm{\boldsymbol{\omega}}||_{1}}{||\mathrm{\boldsymbol{\omega}}||_{2}^{2}}\boldsymbol{\omega}$
was introduced  by Vontobel and Koetter \cite{GraphCover}. They showed
that the decision hyperplane of $\boldsymbol{\omega}$ in the signal
space, is at the same Euclidean distance from $\mathrm{\overline{\mathbf{x}}}_{0}$
and from $\boldsymbol{\overline{\omega}}_{virt}$. Note that if $\boldsymbol{\omega}\in\mathcal{C}\subseteq\{0,1\}^{n}$,
then $\boldsymbol{\omega}_{virt}=\boldsymbol{\omega}$. From the above,
the LP error probability is then expressed in the signal space as
follows.\vspace{-4mm}

\begin{equation}
P_{r}^{LPD}(error\mid\mathrm{\mathbf{x}}_{0})=P_{r}\left\{ \bigcup_{\boldsymbol{\omega}\in\mathcal{G}(\mathcal{\mathcal{K}}(H))}||\overline{\boldsymbol{\omega}}_{virt}-\mathrm{\mathbf{y}}||_{2}\leq||\overline{\mathrm{\mathbf{x}}}_{0}-\mathrm{\mathbf{y}}||_{2}\mid\mathrm{\mathbf{x}}_{0}\right\} .\label{eq:LP error}
\end{equation}
Evaluating the LP error probability by simulating Eq. \eqref{eq:LP error}
is not practical, since it involves enormous number of generators.
However, it allows to make a simulation of the error probability contributed
by a subgroup of generators.

Let $E_{\mathrm{\mathbf{x}}_{0}\rightarrow\boldsymbol{\omega}}=\left\{ \,||\overline{\boldsymbol{\omega}}_{virt}-\mathrm{\mathbf{y}}||_{2}\leq||\overline{\mathrm{\mathbf{x}}}_{0}-\mathrm{\mathbf{y}}||_{2}\mid\mathrm{\mathbf{x}}_{0}\,\right\} $
denote the LP pairwise error event where the received vector $\mathrm{\mathbf{y}}$
is closer to $\overline{\boldsymbol{\omega}}_{virt}$ than to the
transmitted signal $\overline{\mathrm{\mathbf{x}}}_{0}$. Thus the
LP error probability \eqref{eq:LP error} can be written: \vspace{-7mm}

\begin{equation}
P_{r}^{LPD}(error\mid\mathrm{\mathbf{x}}_{0})=P_{r}\left\{ \bigcup_{\boldsymbol{\omega}\in\mathcal{G}(\mathcal{\mathcal{K}}(H))}E_{\mathrm{\mathbf{x}}_{0}\rightarrow\boldsymbol{\omega}}\right\} ,\label{eq:LP errorUn}
\end{equation}
and the LP union bound is \vspace{-8mm}

\begin{equation}
P_{r}^{LPD}(error\mid\mathrm{\mathbf{x}}_{0})\leq\sum_{\boldsymbol{\omega}\in\mathcal{G}(\mathcal{\mathcal{K}}(H))}P_{r}\{E_{\mathrm{\mathbf{x}}_{0}\rightarrow\boldsymbol{\omega}}\}.\label{eq:UB-PW}
\end{equation}
 Let $r_{\boldsymbol{\omega}}\triangleq\tfrac{||\overline{\boldsymbol{\omega}}_{virt}-\overline{\mathrm{\mathbf{x}}}_{0}||_{2}}{2}=\gamma\sqrt{w_{p}(\boldsymbol{\omega})}$
denote the Euclidean distance from $\overline{\mathrm{\mathbf{x}}}_{0}$
or from $\overline{\boldsymbol{\omega}}_{virt}$ to the decision boundary
line. Thus the LP pairwise error probability \cite{GraphCover} \vspace{-7mm}

\begin{equation}
P_{r}(E_{\mathrm{\mathbf{x}}_{0}\rightarrow\boldsymbol{\omega}})=Q\left(\dfrac{r_{\boldsymbol{\omega}}}{\sigma}\right),\label{eq:LPairWise}
\end{equation}
and the LP-UB in Eq. \eqref{eq:UB-PW} can be written as follows \cite{DecisionRegions}
\cite{LP-UB2}. \vspace{-7mm}

\begin{equation}
P_{r}^{LPD}(error\mid\mathrm{\mathbf{x}}_{0})\leq\sum_{\boldsymbol{\omega}\in\mathcal{G}(\mathcal{\mathcal{K}}(H))}Q\left(\dfrac{r_{\boldsymbol{\omega}}}{\sigma}\right).\label{eq:LP-UB}
\end{equation}

\subsection{Undirected Graphs }

In this section, we give a brief overview of some terms from graph
theory. By a graph we will always mean an undirected graph without
loops and multiple edges. We let $|V|$ denote the size of a set $V$.
\begin{defn}
\label{def:UndirectedGraph} (\cite{DataStruct&Alg}) An \emph{undirected
graph} \emph{$G(V,\mathcal{E})$} consists of a set of nodes $V$
and a set of edges $\mathcal{E}$. An edge is an unordered pair of
nodes $(v_{i},v_{j})$. Associated with each edge $(v_{i},v_{j})\in\mathcal{E}$
is a cost $c(v_{i},v_{j})$.
\end{defn}
\vspace{-10mm}\qed
\begin{defn}
(\cite{DataStruct&Alg}) A \emph{spanning tree} of an undirected graph
\emph{$G(V,\mathcal{E}$)}, is a subgraph \emph{$T(V,\mathcal{E}'$)}
that is a tree and connects all the nodes in \emph{$V$}. It has $|V|$
nodes and $|\mathcal{E}'|=|V|-1$ edges, in which \emph{$\mathcal{E}'$}
is a subset of \emph{$\mathcal{E}$. }The cost of a spanning tree
T, denoted by $cost(T)$, is the sum of the costs of all the edges
in the tree. i.e. $cost(T)=\underset{(v_{i},v_{j})\in T}{\sum}c(v_{i},v_{j}).$\qed
\end{defn}

\begin{defn}
(\cite{DataStruct&Alg}) A spanning tree of a graph \emph{$G(V,\mathcal{E})$}
is called a\emph{ Minimum Spanning Tree }(MST), if its cost is less
than or equal to the cost of every other spanning tree \emph{$T(V,\mathcal{E}')$}
of \emph{$G(V,\mathcal{E})$.} \qed
\end{defn}
Two popular algorithms for finding an MST in undirected graph are
Prim's \cite{Prim} and Kruskal's \cite{Kruskal}. A simple implementation
of Prim's algorithm can shows $O(|V|^{2})$ running time, and both
can be implemented with complexity of $O(|\mathcal{E}|\, log|V|)$.

\section{Generator Density Characterization\label{Density} }

In this section, we explore the density of the fundamental cone generators
and we compare it to that of ML codewords. As a result, we will later
examine how the union bound is affected by that density. Let $0\leq\theta_{ij}\leq\pi$
denote the positive angle formed by the vectors $\boldsymbol{\omega}_{i}$
and $\boldsymbol{\omega}_{j}$, which is equal to the angle formed
by the vectors $\,\overrightarrow{\overline{\mathrm{\mathbf{x}}}_{0}\boldsymbol{\overline{\omega}}_{i,}}{}_{virt}$
and $\overrightarrow{\overline{\mathrm{\mathbf{x}}}_{0}\boldsymbol{\overline{\omega}}_{j,}}{}_{virt}$
in a BPSK signal space.

\bigskip{}

\begin{defn}
\label{Def:AngleGraph} Let $\boldsymbol{\omega}_{1},\boldsymbol{\omega}_{2},...,\boldsymbol{\omega}_{M}\in\mathbb{R}_{+}^{n}$
be a set of vectors. Consider each vector as a node of an undirected
graph $G(V,\mathcal{E})$, with an undirected edge joining each pair
of nodes $\boldsymbol{\omega}_{i}$ and $\boldsymbol{\omega}_{j}$,
denoted by $(\boldsymbol{\omega}_{i},\boldsymbol{\omega}_{j})$. An
edge $(\boldsymbol{\omega}_{i},\boldsymbol{\omega}_{j})\in\mathcal{E}$
has a cost that equal to the angle between the vectors related to
the adjacent nodes, i.e, $c(\boldsymbol{\omega}_{i},\boldsymbol{\omega}_{j})=\theta_{ij}$.
The graph $G(V,\mathcal{E})$\emph{ }will be called the\emph{ angle
graph}. Note that the angle graph is a \emph{complete} graph; it has
$|V|$ nodes and $|V|(|V|-1)/2$ edges. \qed
\end{defn}

\begin{figure}[H]
\centering{}\includegraphics[scale=0.14]{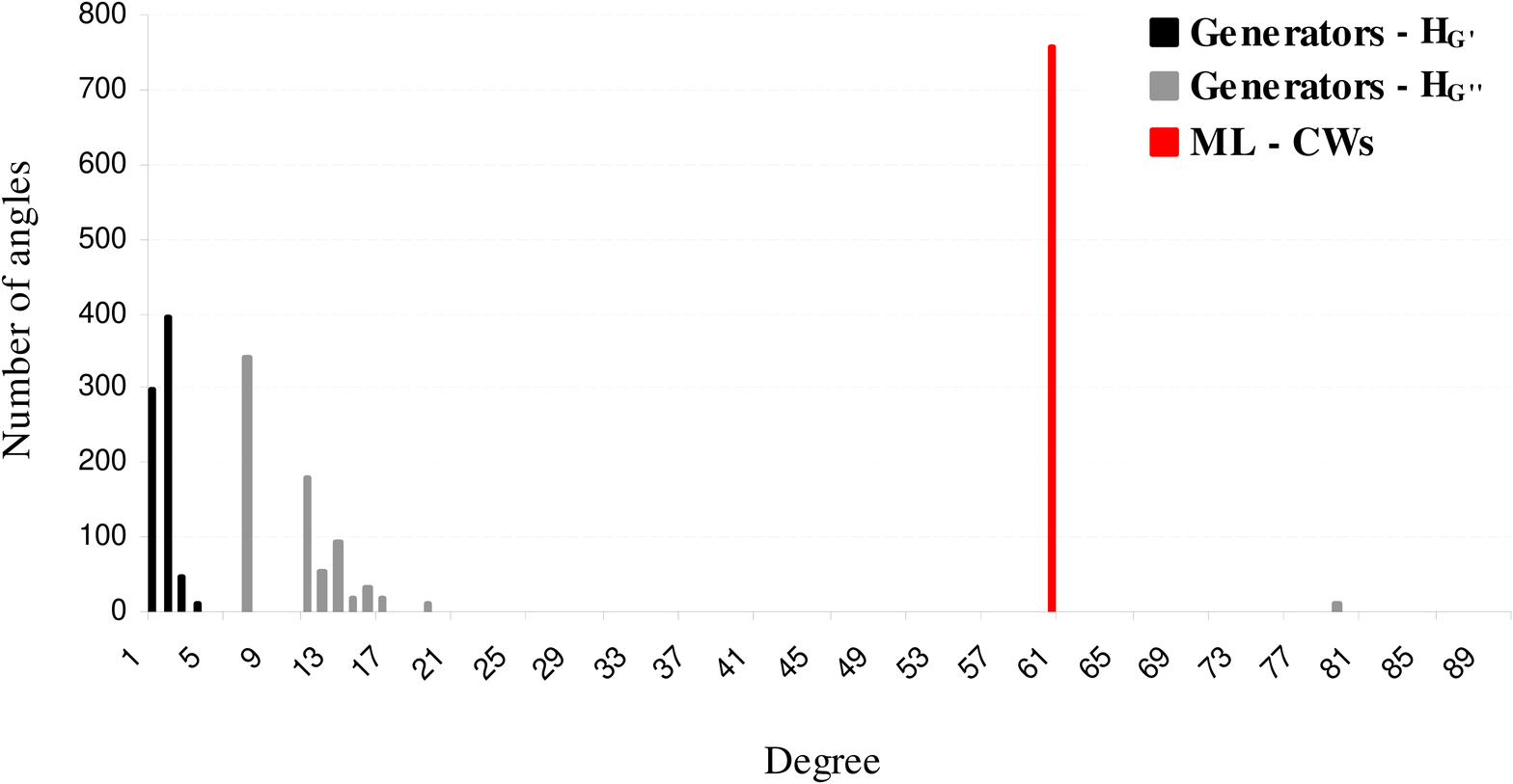} {\footnotesize \caption{{\footnotesize \label{fig:AngleDistribution}}\fontsize{9}{10}\selectfont
Angle distributions for the extended Golay{[}24, 12, 8{]} code of
the first 759 minimal-weight generators of the parity-check matrices
$H_{G'}$ and $H_{G''}$, compared to the angle distribution of the
759 minimal-weight ML codewords.}
}
\end{figure}

\begin{figure}[H]
\centering{}\includegraphics[scale=0.1]{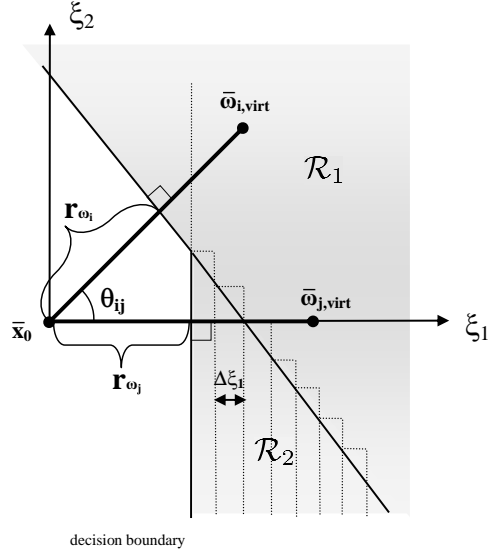} {\scriptsize \caption{{\scriptsize \label{fig:RealcorrectRegion}}{\footnotesize{} }\fontsize{9}{10}\selectfont
The LP triplet-wise error region in the signal space.}
}
\end{figure}

\begin{defn}
\label{def:angle dist.}Let $T(V,\mathcal{E}')$ be an MST of the
angle graph $G(V,\mathcal{E})$ in Def. \ref{Def:AngleGraph}. The\emph{
MST angle distribution} is defined to be the cost distribution of
the all edges $(\boldsymbol{\omega}_{i},\boldsymbol{\omega}_{j})$
in the graph $T(V,\mathcal{E}')$. For easier notation, we will use
the shorter term\emph{ angle distribution} instead. \qed\end{defn}
\begin{example}
\label{Ex:GolayAngDis} Let $H_{G'}$ \cite{DecisionRegions} and
$H_{G''}$ \eqref{eq:HG''} be parity-check matrices for the extended
Golay{[}24,12,8{]} code. The former matrix was introduced by Halford
and Chugg \cite{Golay-H}, the latter is a systematic parity-check
matrix. Fig. \ref{fig:AngleDistribution} presents the angle distributions
of the first 759 minimal-weight generators of $H_{G'}$ and $H_{G''}$
(generators with equal pseudo-weight were ordered randomly.). For
a comparison, the angle distribution of the 759 minimal-weigh ML codewords
is presented as well. The average angle of $H_{G'}$, $H_{G''}$ generators
and of ML codewords are : $1.43^{\circ},$ $10.69^{\circ}$ and $60^{\circ}$,
respectively; and their Standard Deviations (STDs) are: $3.38^{\circ},$
$8.72^{\circ}$ and $0^{\circ}$, respectively. Note that $H_{G'}$
and $H_{G''}$ have two different \emph{generator matrices}, however,
both have the same angle distribution for their $759$ minimal-weight
CWs. It is clear from Fig. \ref{fig:AngleDistribution}, that $H_{G'}$
generators are much crowded than those of $H_{G''}$, and between
these three distributions the ML codewords are spread most widely
and evenly in the Euclidean space.

\begin{spacing}{0.9} 

\setlength\arraycolsep{0.2em} 

\begin{equation}
H_{G''}=\left(\begin{array}{cccccccccccccccccccccccc}
0 & 1 & 1 & 1 & 1 & 1 & 1 & 1 & 1 & 1 & 1 & 1 & 1 & 0 & 0 & 0 & 0 & 0 & 0 & 0 & 0 & 0 & 0 & 0\\
1 & 1 & 1 & 0 & 1 & 1 & 1 & 0 & 0 & 0 & 1 & 0 & 0 & 1 & 0 & 0 & 0 & 0 & 0 & 0 & 0 & 0 & 0 & 0\\
1 & 1 & 0 & 1 & 1 & 1 & 0 & 0 & 0 & 1 & 0 & 1 & 0 & 0 & 1 & 0 & 0 & 0 & 0 & 0 & 0 & 0 & 0 & 0\\
1 & 0 & 1 & 1 & 1 & 0 & 0 & 0 & 1 & 0 & 1 & 1 & 0 & 0 & 0 & 1 & 0 & 0 & 0 & 0 & 0 & 0 & 0 & 0\\
1 & 1 & 1 & 1 & 0 & 0 & 0 & 1 & 0 & 1 & 1 & 0 & 0 & 0 & 0 & 0 & 1 & 0 & 0 & 0 & 0 & 0 & 0 & 0\\
1 & 1 & 1 & 0 & 0 & 0 & 1 & 0 & 1 & 1 & 0 & 1 & 0 & 0 & 0 & 0 & 0 & 1 & 0 & 0 & 0 & 0 & 0 & 0\\
1 & 1 & 0 & 0 & 0 & 1 & 0 & 1 & 1 & 0 & 1 & 1 & 0 & 0 & 0 & 0 & 0 & 0 & 1 & 0 & 0 & 0 & 0 & 0\\
1 & 0 & 0 & 0 & 1 & 0 & 1 & 1 & 0 & 1 & 1 & 1 & 0 & 0 & 0 & 0 & 0 & 0 & 0 & 1 & 0 & 0 & 0 & 0\\
1 & 0 & 0 & 1 & 0 & 1 & 1 & 0 & 1 & 1 & 1 & 0 & 0 & 0 & 0 & 0 & 0 & 0 & 0 & 0 & 1 & 0 & 0 & 0\\
1 & 0 & 1 & 0 & 1 & 1 & 0 & 1 & 1 & 1 & 0 & 0 & 0 & 0 & 0 & 0 & 0 & 0 & 0 & 0 & 0 & 1 & 0 & 0\\
1 & 1 & 0 & 1 & 1 & 0 & 1 & 1 & 1 & 0 & 0 & 0 & 0 & 0 & 0 & 0 & 0 & 0 & 0 & 0 & 0 & 0 & 1 & 0\\
1 & 0 & 1 & 1 & 0 & 1 & 1 & 1 & 0 & 0 & 0 & 1 & 0 & 0 & 0 & 0 & 0 & 0 & 0 & 0 & 0 & 0 & 0 & 1
\end{array}\right)\;\label{eq:HG''}
\end{equation}

\end{spacing} 
\end{example}
 \qed
\begin{example}
\smallskip{}

The error probability contributed by two vectors depends on the angle
between them. Let $\boldsymbol{\omega}_{i},\boldsymbol{\omega}_{j}\in\mathbb{R}_{+}^{n}$
be vectors with an equal pseudo-weight, and let $\xi_{1}$ and $\xi_{2}$
be the two independent Gaussian random variables obtained by projecting
the noise vector $\mathbf{\mathrm{\mathbf{z}}}$ onto the plan determined
by the vectors $\overrightarrow{\overline{\mathrm{\mathbf{x}}}_{0}\boldsymbol{\overline{\omega}}_{i,}}{}_{virt}$
and $\overrightarrow{\overline{\mathrm{\mathbf{x}}}_{0}\boldsymbol{\overline{\omega}}_{j,}}{}_{virt}$.
We refer to the probability $P_{\text{\ensuremath{r}}}\left\{ E_{\mathrm{\mathbf{x}}_{0}\rightarrow\boldsymbol{\omega}_{i}}\bigcup E_{\mathrm{\mathbf{x}}_{0}\rightarrow\boldsymbol{\omega}_{j}}\right\} $
as the triplet-wise error probability, that is $\boldsymbol{\omega}_{i}$
or $\boldsymbol{\omega}_{j}$ was decoded when the all-zeros codeword
was transmitted. The triplet-wise error probability depends on the
angle $\theta_{ij}$ and it can be obtained by integrating a two dimensional
Gaussian distribution over the darkened regions $\mathcal{R}_{1}$
and $\mathcal{R}_{2}$ in Fig. \ref{fig:RealcorrectRegion} \cite{NewTechniques}.
Without loss of generality, we assume that $\boldsymbol{\omega}_{j}$
is placed on $\xi_{1}$ axis. $r_{\boldsymbol{\omega}_{i}}$ and $r_{\boldsymbol{\omega}_{j}}$
denote the Euclidean distances from the decision boundaries lines
of $\boldsymbol{\omega}_{i}$ and $\boldsymbol{\omega}_{j}$, respectively,
to the all-zeros codeword. In the case of vectors of equal pseudo-weight,
$r_{\boldsymbol{\omega}_{i}}=r_{\boldsymbol{\omega}_{j}}$. The decision
region boundary lines of $\boldsymbol{\omega}_{i}$ and $\boldsymbol{\omega}_{j}$
are $\xi_{2}=-a\xi_{1}+b$ and $\xi_{1}=r_{\boldsymbol{\omega}_{j}}$,
respectively. The $\boldsymbol{\omega}_{i}$ boundary line crosses
$\xi_{2}$ axis at point $b=r_{\boldsymbol{\omega}_{i}}/sin\theta_{ij}$
and its slope is $a=\tan(90-\theta_{ij})$. The intersection between
the two boundary lines occurs at point $(\xi'_{1},\xi'_{2})=(r_{\boldsymbol{\omega}_{j}},\:-ar_{\boldsymbol{\omega}_{j}}+b)$. 

There are various numerical integration ways \cite{Numerical} to
evaluate the triplet-wise error probability. Another possibility,
is to approximate it by sum of $Q$-functions as follows.
\end{example}
\smallskip{}

$P_{\text{\ensuremath{r}}}\left\{ E_{\mathrm{\mathbf{x}}_{0}\rightarrow\omega_{i}}\bigcup E_{\mathrm{\mathbf{x}}_{0}\rightarrow\omega_{j}}\right\} =P_{r}\{\mathcal{R}_{1}\}+P_{r}\{\mathcal{R}_{2}\}\approx Q\left(\dfrac{r_{\boldsymbol{\omega}_{i}}}{\sigma}\right)+$

\vspace{-6mm}

\begin{equation}
\overset{\left\lfloor \frac{\xi_{1,max}}{\bigtriangleup\xi_{1}}\right\rfloor }{\underset{k=0}{\sum}}\left[1-Q\left(\frac{-a(\xi'_{1}+k\bigtriangleup\xi_{1})+b}{\sigma}\right)\right]\left[Q\left(\frac{\xi'_{1}+k\bigtriangleup\xi_{1}}{\sigma}\right)-Q\left(\frac{\xi'_{1}+(k+1)\bigtriangleup\xi_{1}}{\sigma}\right)\right].\label{eq:RealCalc}
\end{equation}
$P_{r}\{\mathcal{R}_{1}\}$ is equal to an LP pairwise error probability
\eqref{eq:LPairWise}. $P_{r}\{\mathcal{R}_{2}\}$ is calculated as
follows. The region $\mathcal{R}_{2}$ is divided into rectangles
of a width $\bigtriangleup\xi_{1}$ which are parallel to the $\xi_{2}$
axis, as shown in Fig. \ref{fig:RealcorrectRegion}. Each rectangle
starts from a point on the decision boundary line of $\boldsymbol{\omega}_{i}$
and goes to infinity in the opposite direction of $\xi_{2}$ axis.
The multiplication inside the sum of Eq. \eqref{eq:RealCalc} is the
probability that the noise components $\xi_{1}$ and $\xi_{2}$ are
within the \emph{k}-th rectangle. Since a two dimensional Gaussian
distribution converges to zero as $\xi_{1}$ goes to infinity, it
will be sufficient to sum from $k=0$ to a large $k$ such as $\left\lfloor \frac{\xi_{1,max}}{\bigtriangleup\xi_{1}}\right\rfloor $,
where all the rectangles are located on the left side of the line
$\xi_{2}=\xi_{1,max}$.

\bigskip{}

\begin{example}
\label{exa:RealErrorBCH} Consider the BCH{[}63,57,3{]} code. The
fundamental cone of the systematic parity-check matrix created by
the \emph{generator polynomial} $x^{6}+x+1$ has $11,551$ minimal-weight
generators of pseudo-weight three. The angles between them varied
from $5.85^{\circ}$ to $90^{\circ}$. The triplet-wise error probability
of its two minimal-weight generators depends on $\theta_{ij}$ is
presented in Fig. \ref{fig:UpBVsReal}. It was calculated by Eq. \eqref{eq:RealCalc}
for $0$ and $8$ dB SNR in different angles. The triplet-wise union
bound which is $2Q\left(\tfrac{r_{\boldsymbol{\omega}}}{\sigma}\right)$
is presented as well. $\xi_{1,max}$ and $\bigtriangleup\xi_{1}$
was chosen to be 2000 and 1/2000, respectively. From Fig. \ref{fig:UpBVsReal}
one can observe that the lower the SNR and the smaller the angle are,
the worse is the UB. The figure also presents a triplet-wise error
probability upper bound which is tighter than the UB and it will be
introduced in Sec. \ref{ImprovedUB} 

\begin{figure}[H]
\begin{raggedright}
\begin{minipage}[c][1\totalheight][t]{0.45\textwidth}%
\selectlanguage{british}%
\begin{flushleft}
\subfloat[\selectlanguage{english}%
\fontsize{9}{10}\selectfont SNR = 0 dB\selectlanguage{british}%
]{\selectlanguage{english}%
\includegraphics[scale=0.1]{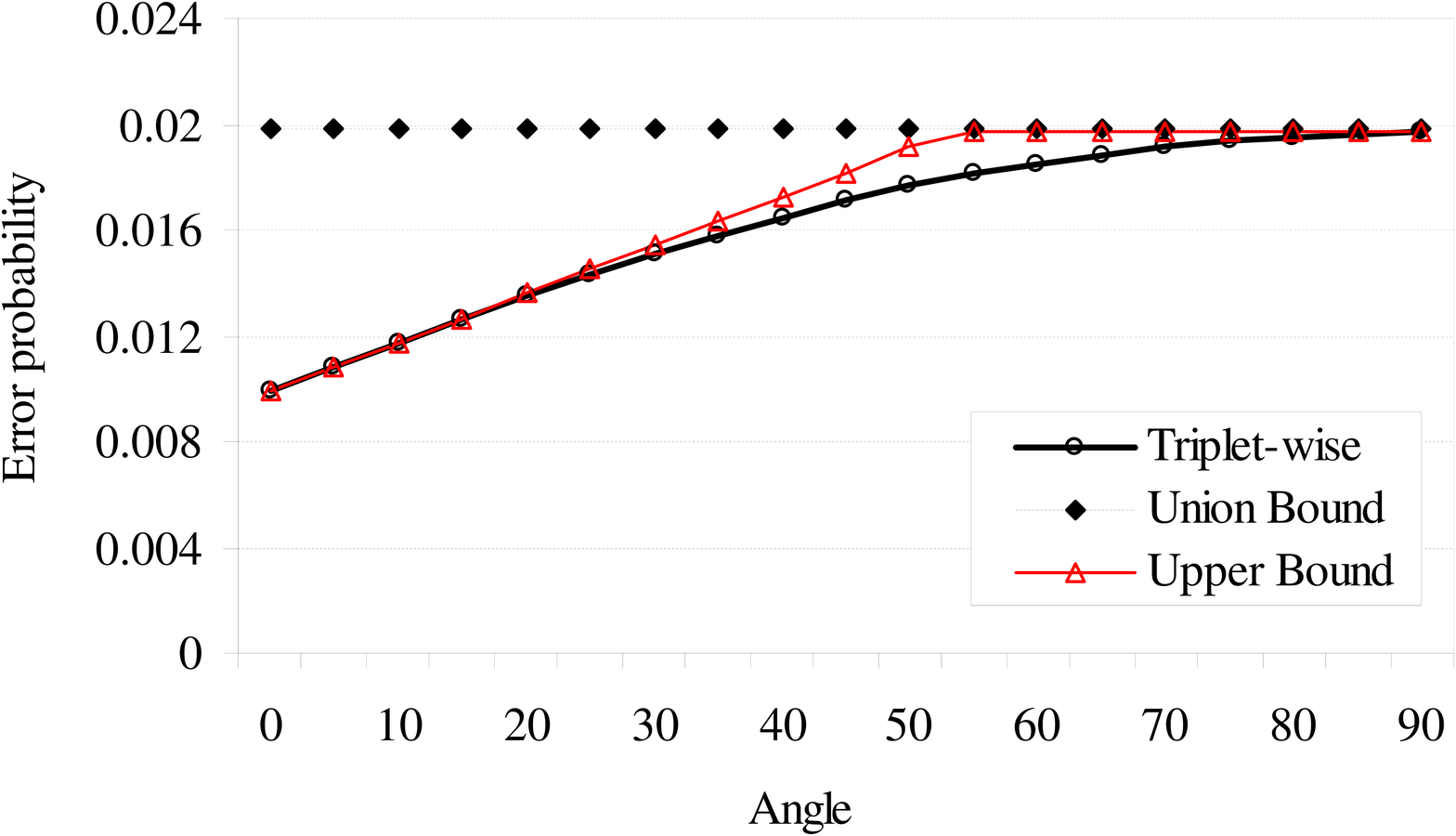}

\selectlanguage{british}%
}
\par\end{flushleft}\selectlanguage{english}%
\end{minipage}\hfill{}%
\begin{minipage}[c][1\totalheight][t]{0.45\textwidth}%
\medskip{}

\selectlanguage{british}%
\begin{flushleft}
\subfloat[\selectlanguage{english}%
\fontsize{9}{10}\selectfont SNR = 8 dB\selectlanguage{british}%
]{\selectlanguage{english}%
\begin{raggedright}
\includegraphics[scale=0.1]{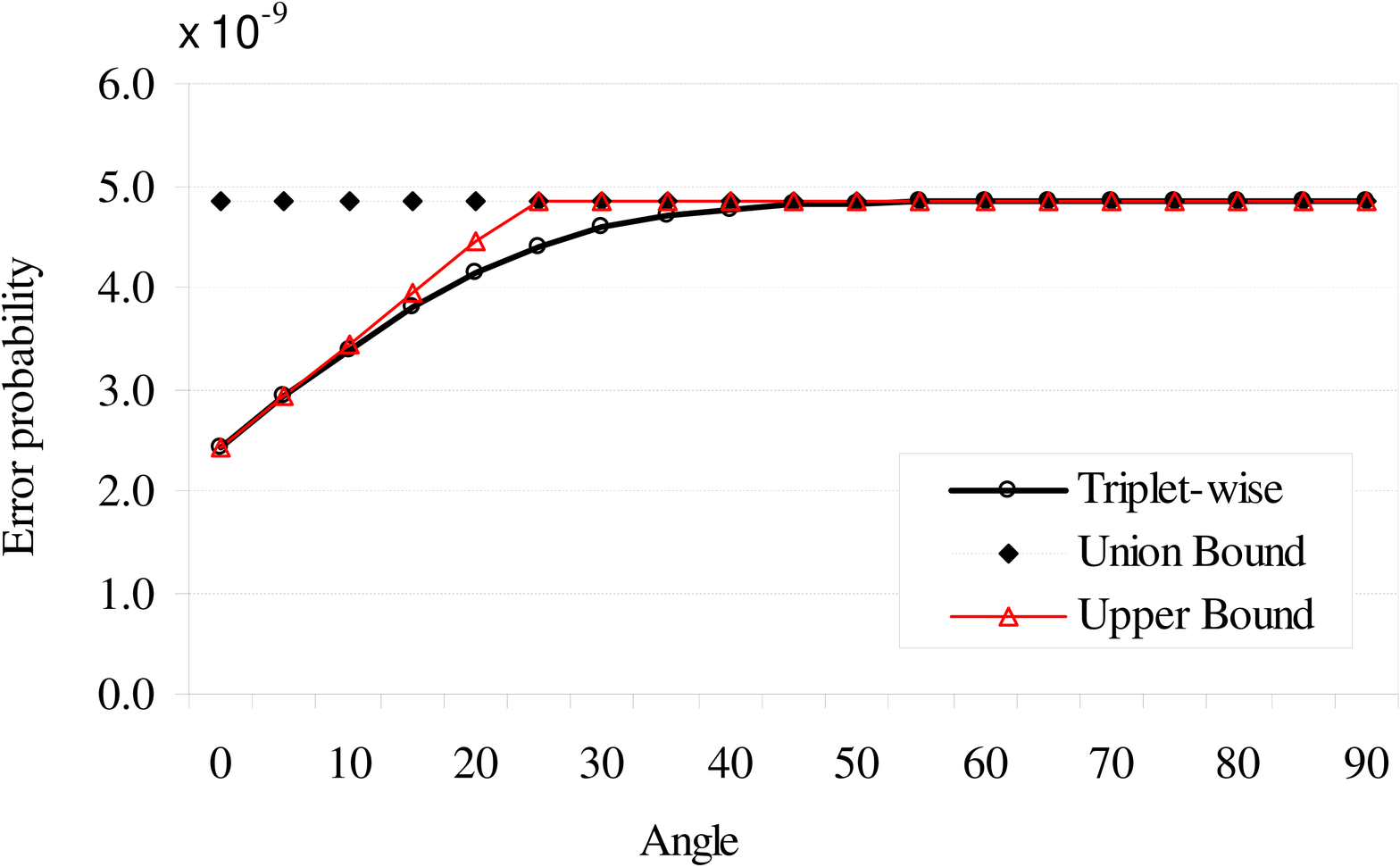}
\par\end{raggedright}

\selectlanguage{british}%
}
\par\end{flushleft}\selectlanguage{english}%
\end{minipage}\hfill{}
\par\end{raggedright}

\caption{\label{fig:UpBVsReal}\fontsize{9}{10}\selectfont Comparison between
the triplet-wise error probability, its union bound and the upper
bound in different angles of two minimal-weight generators of BCH{[}63,57,3{]},
when the all-zeros word was transmitted. }
\end{figure}
 \qed
\end{example}

\section{The Problem of Locating Dominant Error Events of LPD \label{DominantGroup} }

Consider a ML decoding of a binary-linear code BPSK-modulated over
an AWGN channel. The decoder performance can be evaluated by considering
the contributions of the most dominant error events to the probability
of error. That dominant error events, especially in the higher SNR
region, are the minimal weight codewords. 

In this section, we will examine whether the minimal-weight generators
of LP decoding have such a property as well. We let $w_{H}(\mathrm{\mathbf{x}})$\emph{
}denote\emph{ }the\emph{ Hamming weight} of $\mathrm{\mathbf{x}}$,
which is the number of non-zero positions of $\mathrm{\mathbf{x}}$.
Let $w_{H}^{min}(\mathcal{C})$ denote the minimum Hamming weight
of a linear code $\mathcal{C}$, and let $w_{p}^{min}(H)$ denote
the minimum AWGN channel pseudo-weight of a linear code defined by
the parity-check matrix \emph{H}. We will use the shorter notations
$w_{H}^{min}$ and $w_{p}^{min}$ in case where the discussed code
and matrix are mentioned explicitly. We let $\mathrm{\mathcal{K}_{sub}\mathcal{\subset K}}$
denote a sub-cone of the fundamental cone which created by a chosen
subgroup of generators. The LPD$(\mathrm{\mathcal{K}_{sub}})$ Frame
Error Rate (FER) can be obtained by simulating Eq. \eqref{eq:LP error}.
In the next example, we will study the error probability contributed
by a subgroup of codewords and generators for the extended Golay{[}24,12,8{]}
code.
\begin{example}
The extended Golay{[}24,12,8{]} code has a total $4\text{,}096$ codewords
of which 759 have minimal Hamming weight of $w_{H}^{min}=8$. The
fundamental cone of the parity check-matrix $H_{G'}$ has a total
of $231,146,333$ generators of which two have minimal-weight of $w_{p}^{min}=3.6$
\cite{DecisionRegions}. Simulating the error probability by Eq. \eqref{eq:ML error}
shows that the minimal-weight CWs describe well the MLD performance
at the whole range of SNR, which is not the case for the first 759
minimal-weight generators for LPD. For instance, consider the error
rate of $10^{\lyxmathsym{\textminus}2}$, it was found that the difference
between LPD$(\mathrm{\mathcal{K}_{sub}})$ and LPD$(\mathrm{\mathcal{K}})$
is about 2.5 dB. The angle distributions which were presented in Fig.
\ref{fig:AngleDistribution} support this result: the average angle
of that group of generators is as small as $1.43^{\circ}$, and the
average angle of the ML minimal-weight CWs is $60^{\circ}$.\qed
\end{example}
There are number of reasons why the minimal-weight generators are
often not a dominant subgroup of LPD: (a) There is no guarantee for
significant number of generators with minimal pseudo-weight. The fundamental
cone of $H_{G'}$ for example, has only two. (b) A subgroup of generators
can be very crowded, which significantly reduces their contribution
to the error probability. (c) Unlike MLD which has distinct subgroup
of minimal-weight codewords, LPD often has a continuous-like weight
distribution. For example, the BCH{[}31,21,5{]} code of parity-check
matrix $H_{BCH_{[31,21]}}$ \eqref{eq:HBCH31-21 matrix} has 627,052,479
generators. The pseudo-weight distribution of these generators is
presented in Fig. \ref{fig:BCH[31,21] weight}. Its smooth distribution
makes it difficult to locate a minimal-weight dominant subgroup.

In LPD, a potential subgroup to be a dominant is taking all generators
of weight $w_{p}\leq w_{H}^{min}$. This group is not empty since
$w_{p}^{min}\leq w_{H}^{min}$ \cite{PCWs}, however, it may contains
enormous number of generators. For example, Golay{[}24,12,8{]} has
only $759$ minimal-weight CWs of $w_{H}^{min}=8$, but the fundamental
cone of parity-check matrix $H_{G''}$ has $143,757,418$ generators
of weight $w_{p}\leq w_{H}^{min}=8$.

\bigskip{}

\begin{spacing}{1} 

\setlength\arraycolsep{0.1em} 

\begin{equation}
H_{BCH_{[31,21]}}=\left(\begin{array}{ccccccccccccccccccccccccccccccc}
1 & 0 & 0 & 0 & 0 & 0 & 0 & 0 & 0 & 0 & 1 & 1 & 0 & 1 & 0 & 1 & 0 & 1 & 1 & 1 & 1 & 0 & 0 & 1 & 0 & 0 & 1 & 0 & 1 & 0 & 0\\
0 & 1 & 0 & 0 & 0 & 0 & 0 & 0 & 0 & 0 & 0 & 1 & 1 & 0 & 1 & 0 & 1 & 0 & 1 & 1 & 1 & 1 & 0 & 0 & 1 & 0 & 0 & 1 & 0 & 1 & 0\\
0 & 0 & 1 & 0 & 0 & 0 & 0 & 0 & 0 & 0 & 0 & 0 & 1 & 1 & 0 & 1 & 0 & 1 & 0 & 1 & 1 & 1 & 1 & 0 & 0 & 1 & 0 & 0 & 1 & 0 & 1\\
0 & 0 & 0 & 1 & 0 & 0 & 0 & 0 & 0 & 0 & 1 & 1 & 0 & 0 & 1 & 1 & 1 & 1 & 0 & 1 & 0 & 1 & 1 & 0 & 0 & 0 & 0 & 0 & 1 & 1 & 0\\
0 & 0 & 0 & 0 & 1 & 0 & 0 & 0 & 0 & 0 & 0 & 1 & 1 & 0 & 0 & 1 & 1 & 1 & 1 & 0 & 1 & 0 & 1 & 1 & 0 & 0 & 0 & 0 & 0 & 1 & 1\\
0 & 0 & 0 & 0 & 0 & 1 & 0 & 0 & 0 & 0 & 1 & 1 & 1 & 0 & 0 & 1 & 1 & 0 & 0 & 0 & 1 & 1 & 0 & 0 & 1 & 0 & 1 & 0 & 1 & 0 & 1\\
0 & 0 & 0 & 0 & 0 & 0 & 1 & 0 & 0 & 0 & 1 & 0 & 1 & 0 & 0 & 1 & 1 & 0 & 1 & 1 & 1 & 1 & 1 & 1 & 0 & 1 & 1 & 1 & 1 & 1 & 0\\
0 & 0 & 0 & 0 & 0 & 0 & 0 & 1 & 0 & 0 & 0 & 1 & 0 & 1 & 0 & 0 & 1 & 1 & 0 & 1 & 1 & 1 & 1 & 1 & 1 & 0 & 1 & 1 & 1 & 1 & 1\\
0 & 0 & 0 & 0 & 0 & 0 & 0 & 0 & 1 & 0 & 1 & 1 & 1 & 1 & 1 & 1 & 0 & 0 & 0 & 1 & 0 & 1 & 1 & 0 & 1 & 1 & 1 & 1 & 0 & 1 & 1\\
0 & 0 & 0 & 0 & 0 & 0 & 0 & 0 & 0 & 1 & 1 & 0 & 1 & 0 & 1 & 0 & 1 & 1 & 1 & 1 & 0 & 0 & 1 & 0 & 0 & 1 & 0 & 1 & 0 & 0 & 1
\end{array}\right)\qquad\;\label{eq:HBCH31-21 matrix}
\end{equation}

\end{spacing}\vspace{2mm} 

\begin{figure}[H]
\begin{centering}
\includegraphics[scale=0.15]{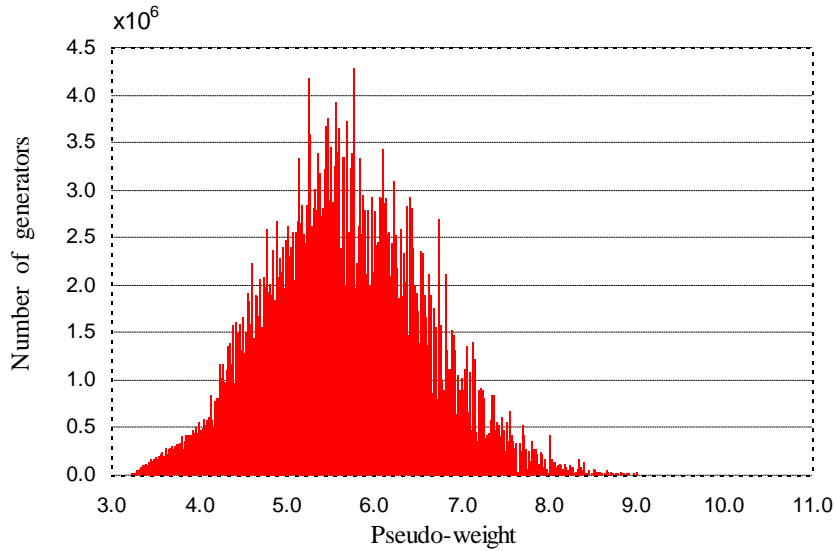}
\par\end{centering}

\caption{{\scriptsize \label{fig:BCH[31,21] weight}}\fontsize{9}{10}\selectfont
A complete generators' pseudo-weight distribution for the BCH{[}31,21,5{]}
code of $H_{BCH_{[31,21]}}$ with 627,052,479 generators.}
\end{figure}

\section{Improved LP Union Bound\label{ImprovedUB} }

In this section, we propose an improved union bound for LP decoding
of a binary linear code transmitted over a binary-input AWGN channel.
This bound is based on the second-order of Bonferroni-type inequality
in probability theory \cite{Bonferroni}, also referred to as Hunter
bound \cite{HunterBound}. For any set of events $E_{1},E_{2},...,E_{M}$
and their \emph{complementary }events, denoted by $E_{1}^{c},E_{2}^{c},...,E_{M}^{c}$,

\begin{equation}
P_{r}\left(\overset{M}{\underset{i=1}{\bigcup}}E_{i}\right)=\overset{M}{\underset{i=1}{\sum}}P_{r}\left(E_{i}\bigcap\left[\,\overset{i-1}{\underset{j=1}{\bigcup}}E_{j}^{c}\right]\right).\label{eq:unionevent}
\end{equation}

\vspace{-1mm} \smallskip{}
Let denote the $M!$ possible permutations of the indices of the error
events $E_{1},E_{2},...,E_{M}$ by $\Pi$(1,2,...,\emph{M}) = $\{\pi_{1},\pi_{2}$,...,$\pi_{M}\}$.
For a given $\Pi$, let $\Lambda=\{\hat{\pi}_{2},\hat{\pi}_{3},...,\hat{\pi}_{M}\}$
denote the $(M^{2}-M)/2$ possible sets of indices in which $\hat{\pi}_{i}\in\{\pi_{1},\pi_{2},...,\pi_{i-1}\}$
for $i=2,3,...,M$. Hunter \cite{HunterBound} presented the second-order
bound of Eq. \eqref{eq:unionevent} as follows.\vspace{-3mm} 

\begin{equation}
P_{r}\left(\overset{M}{\underset{i=1}{\bigcup}}E_{i}\right)\leq\overset{M}{\underset{i=1}{\sum}}P_{r}(E_{\pi_{i}})-\overset{M}{\underset{i=2}{\sum}}P_{r}(E_{\pi_{i}}\cap E_{\hat{\pi}_{i}}).\label{eq:hunter3}
\end{equation}
\vspace{-4mm} 

Minimization of the Right-Hand Side (RHS) of Eq. \eqref{eq:hunter3}
is required to achieve the tightest second-order bound. Using the
sets of the indices $\Lambda$ and $\Pi$, the minimization problem
can be written as follows \cite{NewTSB} \cite{HunterBound}.\vspace{4mm} 

\noindent 
\begin{equation}
P_{r}\left(\overset{M}{\underset{i=1}{\bigcup}}E_{i}\right)\leq\overset{M}{\underset{i=1}{\sum}}P_{r}(E_{i})+\underset{\Pi,\Lambda}{min}\left\{ -\overset{M}{\underset{i=2}{\sum}}P_{r}(E_{\pi_{i}}\cap E_{\hat{\pi}_{i}})\right\} .\label{eq:hunter4}
\end{equation}

\noindent The first sum goes through over all the indices 1 to \emph{M}
of the error events, thus $E_{\pi_{i}}$ could be changed to $E_{i}$.

Consider each of the random events $E_{i}$ as a node of an undirected
graph $G$ and the intersection $(E_{i}\cap E_{j})$ as an undirected
edge joining the nodes $E_{i}$ and $E_{j}$, denoted by $(i,j)$,
with a cost $c(i,j)=P_{r}(E_{i}\cap E_{j})$. Hunter \cite{HunterBound}
showed that a set of $(M-1)$ intersections may be used in the second
term of Eq. \eqref{eq:hunter4} if and only if it forms a spanning
tree of the nodes $\left\{ E_{i}\right\} _{i=1}^{M}$. Thus the minimization
problem of Eq. \eqref{eq:hunter4} can be written equivalently \cite{HunterBound},
\cite{NewTSB}, \vspace{-2mm} 

\begin{equation}
P_{r}\left(\overset{M}{\underset{i=1}{\bigcup}}E_{i}\right)\leq\overset{M}{\underset{i=1}{\sum}}P_{r}(E_{i})+\underset{\tau}{min}\left\{ -\overset{}{\underset{(i,j)\in\tau}{\sum}}P_{r}(E_{i}\cap E_{j})\right\} .\label{eq:hunter5}
\end{equation}

\noindent Where $\tau$ is a spanning tree of the graph $G$. The
problem is to find a graph $\tau$ which minimizes Eq. \eqref{eq:hunter5}
over all possible spanning trees. The solution for that is known as
the solution of the minimum spanning tree problem and has been proposed
by Prim \cite{Prim} and Kruskal \cite{Kruskal}. 

\bigskip{}

Consider the event $E_{i}$ as the pairwise error event $E_{\mathrm{\mathbf{x}}_{0}\rightarrow\boldsymbol{\omega}_{i}}$.
In order to upper bound the LP decoding error probability in Eq. \eqref{eq:LP errorUn}
by the second-order upper bound \eqref{eq:hunter5}, the probability
$P_{\text{\ensuremath{r}}}\left\{ E_{\mathrm{\mathbf{x}}_{0}\rightarrow\boldsymbol{\omega}_{i}}\bigcap E_{\mathrm{\mathbf{x}}_{0}\rightarrow\boldsymbol{\omega}_{j}}\right\} $
is required, or instead, its lower bound. The probability of intersection
of two events can be expressed using the \emph{inclusion-exclusion}
principle in probability theory, \vspace{-7mm} 

\begin{equation}
P_{\text{\ensuremath{r}}}\left\{ E_{\mathrm{\mathbf{x}}_{0}\rightarrow\boldsymbol{\omega}_{i}}\bigcap E_{\mathrm{\mathbf{x}}_{0}\rightarrow\boldsymbol{\omega}_{j}}\right\} =P_{\text{\ensuremath{r}}}\left\{ E_{\mathrm{\mathbf{x}}_{0}\rightarrow\boldsymbol{\omega}_{i}}\right\} +P_{\text{\ensuremath{r}}}\left\{ E_{\mathrm{\mathbf{x}}_{0}\rightarrow\boldsymbol{\omega}_{j}}\right\} -P_{\text{\ensuremath{r}}}\left\{ E_{\mathrm{\mathbf{x}_{0}}\rightarrow\boldsymbol{\omega}_{i}}\bigcup E_{\mathrm{\mathbf{x}}_{0}\rightarrow\boldsymbol{\omega}_{j}}\right\} .\label{eq:Intersect}
\end{equation}
The first and the second terms in the RHS of Eq. \eqref{eq:Intersect}
are the LP pairwise error probability \eqref{eq:LPairWise}, the third
term can be upper bounded by the following theorem.

\bigskip{}

\begin{thm}
\label{thm:UB TWE} Let $\boldsymbol{\omega}_{i},\boldsymbol{\omega}_{j}\in\mathbb{R}_{+}^{n}$
be vectors of a pseudo-weight $w_{p}(\boldsymbol{\omega}_{i})\neq w_{p}(\boldsymbol{\omega}_{j})$.
The LP triplet-wise error probability \vspace{-8mm}
\end{thm}
\begin{equation}
P_{\text{\ensuremath{r}}}\left\{ E_{\mathrm{\mathbf{x}_{0}}\rightarrow\boldsymbol{\omega}_{i}}\bigcup E_{\mathrm{\mathbf{x}}_{0}\rightarrow\boldsymbol{\omega}_{j}}\right\} \leq min\left\{ \begin{array}{c}
Q\left(\dfrac{min(r_{\boldsymbol{\omega}_{i}},r_{\boldsymbol{\omega}_{j}})}{\sigma}\right)+\dfrac{\theta_{ij}}{2\pi}e^{-\frac{max(r_{\boldsymbol{\omega}_{i}}^{2},r_{\boldsymbol{\omega}_{j}}^{2})}{2\sigma^{2}}},\\
Q\left(\dfrac{r_{\boldsymbol{\omega}_{i}}}{\sigma}\right)+Q\left(\dfrac{r_{\boldsymbol{\omega}_{j}}}{\sigma}\right)-Q\left(\dfrac{r_{\boldsymbol{\omega}_{i}}}{\sigma}\right)Q\left(\dfrac{r_{\boldsymbol{\omega}_{j}}}{\sigma}\right)
\end{array}\right\} .\label{eq:TriUpperBd}
\end{equation}

\bigskip{}

\begin{figure}[H]
\centering{}\includegraphics[scale=0.1]{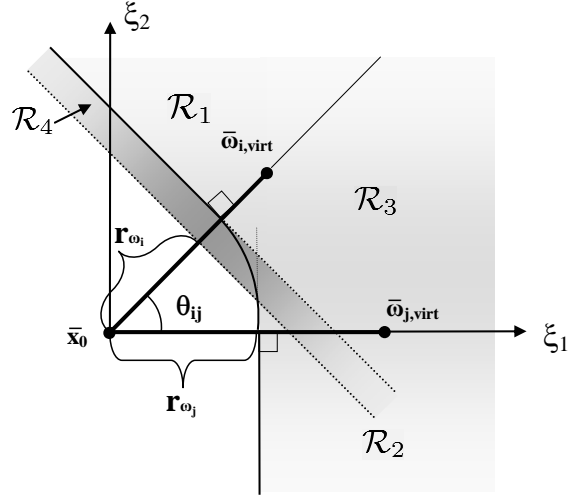} {\footnotesize \caption{\label{fig:Error-probability-area diff}\fontsize{9}{10}\selectfont
The region in the signal space used to bound the LP triplet-wise error
probability ($w_{p}(\boldsymbol{\omega}_{i})\neq w_{p}(\boldsymbol{\omega}_{j})).$}
}
\end{figure}

\begin{IEEEproof}
Let $\tilde{\xi}\triangleq\xi_{1}^{2}+\xi_{2}^{2}$ be a random variable
with Chi-square distribution \cite{Chai} with two degrees of freedom,
i.e. \vspace{-8mm}

\begin{equation}
f(\tilde{\xi})=\frac{1}{2\sigma^{2}}e^{-\frac{\tilde{\xi}}{2\sigma^{2}}}U(\tilde{\xi}),\label{eq:ChiDis}
\end{equation}
in which $U(\cdot)$ is the unit step function. Without loss of generality
we assume that $w_{p}(\boldsymbol{\omega}_{i})<w_{p}(\boldsymbol{\omega}_{j})$.
With the help of Fig. \ref{fig:Error-probability-area diff} the triplet-wise
error probability,\vspace{-7mm}

\begin{eqnarray}
P_{\text{\ensuremath{r}}}\left\{ E_{\mathrm{\mathbf{x}_{0}}\rightarrow\boldsymbol{\omega}_{i}}\bigcup E_{\mathrm{\mathbf{x}}_{0}\rightarrow\boldsymbol{\omega}_{j}}\right\}  & \leq & P_{r}\left(\overset{4}{\underset{i=1}{\bigcup}}\mathcal{R}_{i}\right)\leq\overset{4}{\underset{i=1}{\sum}}P_{r}\{\mathcal{R}_{i}\}\qquad\qquad\qquad\qquad\qquad\qquad\qquad\quad\\
 & = & \underset{P_{r}(\mathcal{R}_{1})+P_{r}(\mathcal{R}_{2})}{\underbrace{Q\left(\dfrac{r_{\boldsymbol{\omega}_{j}}}{\sigma}\right)}}+\underset{P_{r}(\mathcal{R}_{3})}{\underbrace{\frac{\theta_{ij}}{2\pi}P_{r}\left(\tilde{\xi}>r_{\boldsymbol{\omega}_{j}}^{2}\right)}}+\underset{P_{r}(\mathcal{R}_{4})}{\underbrace{Q\left(\dfrac{r_{\boldsymbol{\omega}_{i}}}{\sigma}\right)-Q\left(\dfrac{r_{\boldsymbol{\omega}_{j}}}{\sigma}\right)}}\label{eq:UsingChi}\\
 & = & Q\left(\dfrac{r_{\boldsymbol{\omega}_{i}}}{\sigma}\right)+\dfrac{\theta_{ij}}{2\pi}e^{-\frac{r_{\boldsymbol{\omega}_{j}}^{2}}{2\sigma^{2}}}.
\end{eqnarray}

\noindent \medskip{}
From the noise symmetry, each of the probabilities $P_{r}(\mathcal{R}_{1})$
or $P_{r}(\mathcal{R}_{2})$ equal to $\tfrac{1}{2}Q\left(\tfrac{r_{\boldsymbol{\omega}_{j}}}{\sigma}\right)$.
$P_{r}(\mathcal{R}_{3})$ is the probability that of $\xi_{1}^{2}+\xi_{2}^{2}$
lies in the region outside a circle of a radios $r_{\omega_{j}}$
created by the central angle $\theta_{ij}$. $P_{r}\left(\tilde{\xi}>r_{\boldsymbol{\omega}_{j}}^{2}\right)$
was calculated in Eq. \eqref{eq:UsingChi} by integrating the Chi-square
distribution \eqref{eq:ChiDis} from $r_{\boldsymbol{\omega}_{j}}^{2}$
to $\infty$. Thus for two vectors of pseudo-weight $w_{p}(\boldsymbol{\omega}_{i})\neq w_{p}(\boldsymbol{\omega}_{j})$
\vspace{-4mm}

\begin{equation}
P_{\text{\ensuremath{r}}}\left\{ E_{\mathrm{\mathbf{x}_{0}}\rightarrow\boldsymbol{\omega}_{i}}\bigcup E_{\mathrm{\mathbf{x}}_{0}\rightarrow\boldsymbol{\omega}_{j}}\right\} \leq Q\left(\dfrac{min(r_{\boldsymbol{\omega}_{i}},r_{\boldsymbol{\omega}_{j}})}{\sigma}\right)+\dfrac{\theta_{ij}}{2\pi}e^{-\frac{max(r_{\boldsymbol{\omega}_{i}}^{2},r_{\boldsymbol{\omega}_{j}}^{2})}{2\sigma^{2}}}.\label{eq:BoundUniondiffEvent}
\end{equation}

\begin{flushleft}
\smallskip{}
 The triplet-wise error probability can also be bounded using the
inclusion\textendash{}exclusion principle as follows. \vspace{-6mm} 
\par\end{flushleft}

\begin{spacing}{1} 

\begin{eqnarray}
P_{\text{\ensuremath{r}}}\left\{ E_{\mathrm{\mathbf{x}_{0}}\rightarrow\boldsymbol{\omega}_{i}}\bigcup E_{\mathrm{\mathbf{x}}_{0}\rightarrow\boldsymbol{\omega}_{j}}\right\}  & = & P_{\text{\ensuremath{r}}}\left\{ E_{\mathrm{\mathbf{x}}_{0}\rightarrow\boldsymbol{\omega}_{i}}\right\} +P_{\text{\ensuremath{r}}}\left\{ E_{\mathrm{\mathbf{x}}_{0}\rightarrow\boldsymbol{\omega}_{j}}\right\} -P_{\text{\ensuremath{r}}}\left\{ E_{\mathrm{\mathbf{x}}_{0}\rightarrow\boldsymbol{\omega}_{i}}\bigcap E_{\mathrm{\mathbf{x}}_{0}\rightarrow\boldsymbol{\omega}_{j}}\right\} \label{eq:exl-incl}\\
\nonumber \\
 & \leq & Q\left(\dfrac{r_{\boldsymbol{\omega}_{i}}}{\sigma}\right)+Q\left(\dfrac{r_{\boldsymbol{\omega}_{j}}}{\sigma}\right)-Q\left(\dfrac{r_{\boldsymbol{\omega}_{i}}}{\sigma}\right)Q\left(\dfrac{r_{\boldsymbol{\omega}_{j}}}{\sigma}\right).\label{eq:90d bound-d}
\end{eqnarray}

\end{spacing}

\vspace{5mm} 

\noindent The transition from Eq. \eqref{eq:exl-incl} to Eq. \eqref{eq:90d bound-d}
was done by lower bounding $P_{\text{\ensuremath{r}}}\left\{ E_{\mathrm{\mathbf{x}}_{0}\rightarrow\boldsymbol{\omega}_{i}}\bigcap E_{\mathrm{\mathbf{x}}_{0}\rightarrow\boldsymbol{\omega}_{j}}\right\} $
at its lowest value $Q\left(\tfrac{r_{\boldsymbol{\omega}_{i_{\,}}}}{\sigma}\right)Q\left(\tfrac{r_{\boldsymbol{\omega}_{j}}}{\sigma}\right)$
accepted in $\theta_{ij}=90^{0}$. Finally, selecting the minimum
between Eq. \eqref{eq:BoundUniondiffEvent} and Eq. \eqref{eq:90d bound-d}
completes the proof. 
\end{IEEEproof}

\begin{example}
We continue Ex. \ref{exa:RealErrorBCH}. The triplet-wise error probability
upper bound of Theorem \ref{thm:UB TWE} was calculated for two minimal-weight
generators of the BCH{[}63,57,3{]} code. It is presented in Fig. \ref{fig:UpBVsReal}
together with the previous results of Ex. \ref{exa:RealErrorBCH}.
We can see that the smaller the angle and lower the SNR, the more
improvement the triplet-wise error upper bound has over the union
bound. Note that because  $\tfrac{r_{\boldsymbol{\omega}}}{\sigma}\propto\sqrt{\mathrm{SNR\cdot}w_{p}(\boldsymbol{\omega})}$,
changing the pseudo-weight of the generators will have the same effect
as changing the SNR. Thus this bound is expected to have more improvement
on low pseudo-weight generators. \qed
\end{example}
\noindent In the next theorem, we propose an improved UB for the LP
decoding.
\begin{thm}
\label{thm:ILP-UB} Let $\mathcal{G}(\mathcal{K}(H))$ be a set of
cone generators of a parity-check matrix H. For each \textup{$\boldsymbol{\omega}_{i}\in\mathcal{G}$}
the pairwise error event $E_{\mathrm{\mathbf{x}}_{0}\rightarrow\boldsymbol{\omega}_{i}}$
is considered as a node of a complete graph $G(V,\mathcal{E})$. Let
$(\boldsymbol{\omega}_{i},\boldsymbol{\omega}_{j})$ denote an undirected
edge joining the nodes related to the events $E_{\mathrm{\mathbf{x}}_{0}\rightarrow\boldsymbol{\omega}_{i}}$
and $E_{\mathrm{\mathbf{x}}_{0}\rightarrow\boldsymbol{\omega}_{j}}$.
$\tau(V,\mathcal{E}')$ is denoted for a spanning tree of $G(V,\mathcal{E})$.
The LP decoding error probability can be upper-bounded by
\end{thm}
\setlength\arraycolsep{0.05em} \vspace{-12mm}
\begin{eqnarray}
P_{r}^{LPD}(error\mid\mathrm{\mathbf{x}}_{0}) & \leq & \overset{}{\underset{\boldsymbol{\omega}\in\mathcal{G}(\mathcal{\mathcal{K}}(H))}{\sum}}Q\left(\dfrac{r_{\boldsymbol{\omega}}}{\sigma}\right)\nonumber \\
 & + & \underset{\tau}{min}\left\{ \overset{}{\underset{(\boldsymbol{\omega}_{i},\boldsymbol{\omega}_{j})\in\tau}{\sum}}min\left\{ \begin{array}{c}
-Q\left(\dfrac{max(r_{\boldsymbol{\omega}_{i}},r_{\boldsymbol{\omega}_{j}})}{\sigma}\right)+\dfrac{\theta_{ij}}{2\pi}e^{-\frac{max(r_{\boldsymbol{\omega}_{i}}^{2},r_{\boldsymbol{\omega}_{j}}^{2})}{2\sigma^{2}}},\\
-Q\left(\dfrac{r_{\boldsymbol{\omega}_{i}}}{\sigma}\right)Q\left(\dfrac{r_{\boldsymbol{\omega}_{j}}}{\sigma}\right)
\end{array}\right\} \right\} \:\label{eq:ILP-UB}
\end{eqnarray}
\\
We call this bound the \emph{Improved LP Union Bound} (ILP-UB). The
first term is the LP union bound itself \eqref{eq:LP-UB}, the second
term is a second-order correction. 
\begin{IEEEproof}
To prove this, we will apply Hunter bound for the LP error probability.
First, we find a lower bound for $P_{\text{\ensuremath{r}}}\left\{ E_{\mathrm{\mathbf{x}}_{0}\rightarrow\boldsymbol{\omega}_{i}}\bigcap E_{\mathrm{\mathbf{x}}_{0}\rightarrow\boldsymbol{\omega}_{j}}\right\} $:
by substituting the upper bound of $P_{\text{\ensuremath{r}}}\left\{ E_{\mathrm{\mathbf{x}_{0}}\rightarrow\boldsymbol{\omega}_{i}}\bigcup E_{\mathrm{\mathbf{x}}_{0}\rightarrow\boldsymbol{\omega}_{j}}\right\} $
\eqref{eq:TriUpperBd} into the inclusion\textendash{}exclusion principal
\eqref{eq:Intersect}, we will have

\begin{flushleft}
\smallskip{}
$P_{\text{\ensuremath{r}}}\left\{ E_{\mathrm{\mathbf{x}}_{0}\rightarrow\boldsymbol{\omega}_{i}}\bigcap E_{\mathrm{\mathbf{x}}_{0}\rightarrow\boldsymbol{\omega}_{j}}\right\} \geq$\vspace{-10mm}
\par\end{flushleft}

\begin{eqnarray}
 & \geq & Q\left(\dfrac{r_{\boldsymbol{\omega}_{i}}}{\sigma}\right)+Q\left(\dfrac{r_{\boldsymbol{\omega}_{j}}}{\sigma}\right)-min\left\{ \begin{array}{c}
Q\left(\dfrac{min(r_{\boldsymbol{\omega}_{i}},r_{\boldsymbol{\omega}_{j}})}{\sigma}\right)+\dfrac{\theta_{ij}}{2\pi}e^{-\frac{max(r_{\boldsymbol{\omega}_{i}}^{2},r_{\boldsymbol{\omega}_{j}}^{2})}{2\sigma^{2}}},\\
Q\left(\dfrac{r_{\boldsymbol{\omega}_{i}}}{\sigma}\right)+Q\left(\dfrac{r_{\boldsymbol{\omega}_{j}}}{\sigma}\right)-Q\left(\dfrac{r_{\boldsymbol{\omega}_{i}}}{\sigma}\right)Q\left(\dfrac{r_{\boldsymbol{\omega}_{j}}}{\sigma}\right)
\end{array}\right\} \\
 &  & \tfrac{}{}\nonumber \\
 & = & max\left\{ \begin{array}{c}
Q\left(\dfrac{max(r_{\boldsymbol{\omega}_{i}},r_{\boldsymbol{\omega}_{j}})}{\sigma}\right)-\dfrac{\theta_{ij}}{2\pi}e^{-\frac{max(r_{\boldsymbol{\omega}_{i}}^{2},r_{\boldsymbol{\omega}_{j}}^{2})}{2\sigma^{2}}},\\
Q\left(\dfrac{r_{\boldsymbol{\omega}_{i}}}{\sigma}\right)Q\left(\dfrac{r_{\boldsymbol{\omega}_{j}}}{\sigma}\right)
\end{array}\right\} .\label{eq:InterLower}
\end{eqnarray}

\noindent \smallskip{}

\noindent Applying Hunter bound \eqref{eq:hunter5} for LP decoding
error probability \eqref{eq:LP errorUn} and substituting into it
the expression in \eqref{eq:InterLower} together with the LP pairwise
error probability \eqref{eq:LPairWise}, will give the desired result.
\end{IEEEproof}
Given a set of generators $\mathcal{G}$, the running time of ILP-UB
is equal to that of finding an MST on a complete graph $G(V,\mathcal{E})$.
It can be obtained by Prim's algorithm with a complexity of $O(|\mathcal{G}|^{2})$.
The LP-UB for a comparison, for a given set of generators has running
time of $O(|\mathcal{G}|)$.

\section{Results and Discussion\label{Results} }

In this section, we provide results to show the improvement of ILP-UB
over LP-UB. For this purpose, we examine four codes, three HDPC codes:
extended Golay{[}24,12,8{]}, BCH{[}31,26,3{]}, BCH{[}63,57,3{]}; and
one LDPC Tanner code {[}155,64,20{]} \cite{Tanner155}. The parity-check
matrices we use for Golay{[}24,12,8{]} and BCH{[}31,26,3{]} are $H_{G}''$
\eqref{eq:HG''} and $H_{BCH_{[31,26]}}$ \eqref{eq:HBCH31-26 matrix},
respectively; and for the BCH{[}63,57,3{]} we use a systematic parity-check
matrix created by the generator polynomial $x^{6}+x+1$. The minimal
pseudo-weight of the extended Golay{[}24,12,8{]} is $w_{p}^{min}=3.2$.
BCH{[}31,26,3{]} and BCH{[}63,57,3{]} have the same minimal pseudo-weight:
$w_{p}^{min}=3$; and the Tanner code {[}155,64,20{]} has $w_{p}^{min}\approx16.403$
\cite{DecisionRegions}. 

Because of the enormous number of cone generators, we chose representative
subgroups: for the BCH{[}31,26,3{]}, BCH{[}63,57,3{]} and Tanner code
{[}155,64,20{]} we chose all the minimal-weight generators that are
1,185 , 11,551 and 465 generators, respectively. Because the extended
Golay{[}24,12,8{]} code has only 165 minimal-weight generators we
chose for it the first 231 generators of a weight equal or less than
$w_{p}=3.25$.

\bigskip{}

\begin{spacing}{1} 

\setlength\arraycolsep{0.1em} 

\begin{equation}
H_{BCH_{[31,26]}}=\left(\begin{array}{ccccccccccccccccccccccccccccccc}
1 & 0 & 0 & 0 & 0 & 1 & 0 & 0 & 1 & 0 & 1 & 1 & 0 & 0 & 1 & 1 & 1 & 1 & 1 & 0 & 0 & 0 & 1 & 1 & 0 & 1 & 1 & 1 & 0 & 1 & 0\\
0 & 1 & 0 & 0 & 0 & 0 & 1 & 0 & 0 & 1 & 0 & 1 & 1 & 0 & 0 & 1 & 1 & 1 & 1 & 1 & 0 & 0 & 0 & 1 & 1 & 0 & 1 & 1 & 1 & 0 & 1\\
0 & 0 & 1 & 0 & 0 & 1 & 0 & 1 & 1 & 0 & 0 & 1 & 1 & 1 & 1 & 1 & 0 & 0 & 0 & 1 & 1 & 0 & 1 & 1 & 1 & 0 & 1 & 0 & 1 & 0 & 0\\
0 & 0 & 0 & 1 & 0 & 0 & 1 & 0 & 1 & 1 & 0 & 0 & 1 & 1 & 1 & 1 & 1 & 0 & 0 & 0 & 1 & 1 & 0 & 1 & 1 & 1 & 0 & 1 & 0 & 1 & 0\\
0 & 0 & 0 & 0 & 1 & 0 & 0 & 1 & 0 & 1 & 1 & 0 & 0 & 1 & 1 & 1 & 1 & 1 & 0 & 0 & 0 & 1 & 1 & 0 & 1 & 1 & 1 & 0 & 1 & 0 & 1
\end{array}\right)\qquad\;\;\;\label{eq:HBCH31-26 matrix}
\end{equation}

\end{spacing}

\bigskip{}

\bigskip{}
Fig. \ref{fig:ResultesAngledistribution} presents the angle distributions
according to Def. \ref{def:angle dist.} for the aforementioned codes:
extended Golay{[}24,12,8{]}, BCH{[}31,26,3{]} and BCH{[}63,57,3{]}.
Their average angles are $19.85^{\circ}$, $29.58^{\circ}$, $21.87^{\circ}$,
respectively; and their STDs are $13.44^{\circ}$, $13.94^{\circ}$,
$13.84^{\circ}$, respectively.

\begin{flushleft}
\begin{figure}[H]
\begin{raggedleft}
\begin{minipage}[c][1\totalheight][t]{0.45\textwidth}%
\selectlanguage{british}%
\begin{flushleft}
\subfloat[{\selectlanguage{english}%
{\scriptsize \label{fig:HG'' AnglesDis}}\fontsize{9}{10}\selectfont
Golay{[}24,12,8{]} code: angle distribution for all the $231$ generators
with $w_{p}\leq3.25$.\selectlanguage{british}%
}]{\selectlanguage{english}%
\begin{centering}
\includegraphics[scale=0.1]{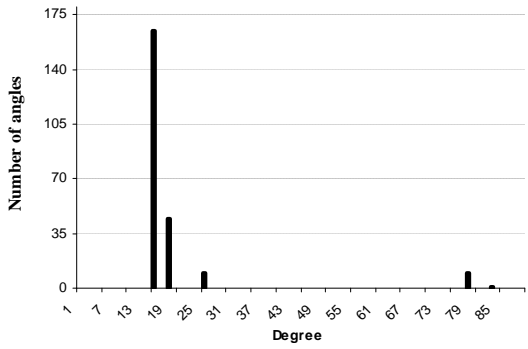}
\par\end{centering}

\selectlanguage{british}%
}
\par\end{flushleft}\selectlanguage{english}%
\end{minipage}\hfill{}%
\begin{minipage}[c][1\totalheight][t]{0.45\textwidth}%
\medskip{}

\selectlanguage{british}%
\begin{flushright}
\subfloat[{\selectlanguage{english}%
{\scriptsize \label{fig:BCH(31,26) AnglesDis}}\fontsize{9}{10}\selectfont
BCH{[}31,26,3{]} code: angle distribution of all the 1,185 minimal-weight
generators.\selectlanguage{british}%
}]{\selectlanguage{english}%
\begin{centering}
\includegraphics[scale=0.1]{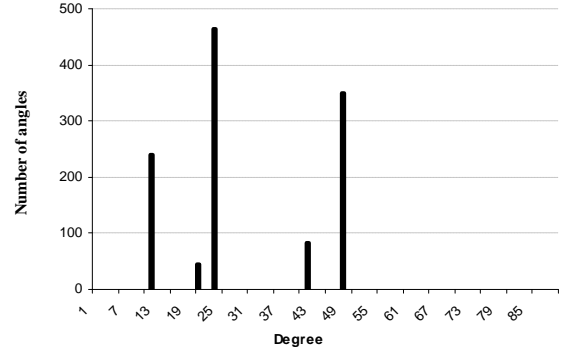}
\par\end{centering}

\selectlanguage{british}%
}
\par\end{flushright}\selectlanguage{english}%
\end{minipage}\hfill{}
\par\end{raggedleft}

\begin{centering}
\begin{minipage}[c][1\totalheight][t]{0.45\textwidth}%
\bigskip{}

\selectlanguage{british}%
\begin{flushleft}
\subfloat[{\selectlanguage{english}%
{\scriptsize \label{fig:BCH(63,57) anglesDis}}\fontsize{9}{10}\selectfont
BCH{[}63,57,3{]} code: angle distribution of all the 11,551 minimal-weight
generators.\selectlanguage{british}%
}]{\selectlanguage{english}%
\begin{centering}
\includegraphics[scale=0.1]{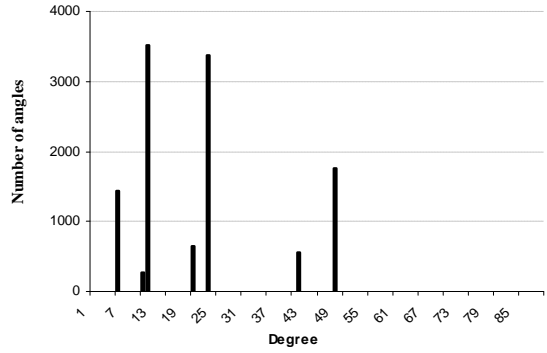}
\par\end{centering}

\selectlanguage{british}%
}
\par\end{flushleft}\selectlanguage{english}%
\end{minipage}
\par\end{centering}

\caption{\label{fig:ResultesAngledistribution}\fontsize{9}{10}\selectfont
Angle distributions.}
\end{figure}

\par\end{flushleft}

Fig. \ref{fig::ResultesILP-UB} presents results of the: ILP-UB$(\mathrm{\mathcal{K}_{sub}})$,
LP-UB$(\mathrm{\mathcal{K}_{sub}})$ and LPD$(\mathrm{\mathcal{K}_{sub}})$
for the chosen subgroups of generators. It presents the LPD FER as
well. The ILP-UB optimized by Prim's algorithm. The ILP-UB presents
an improvement over the LP-UB. For instance, we consider the error
rate of $10^{\lyxmathsym{\textminus}2}$. For the extended Golay{[}24,12,8{]},
the difference between LP-UB$(\mathrm{\mathcal{K}_{sub}})$ and LPD$(\mathrm{\mathcal{K}_{sub}})$
is about 0.9 dB while ILP-UB$(\mathrm{\mathcal{K}_{sub}})$ shows
an improvement of 0.37 dB over LP-UB$(\mathrm{\mathcal{K}_{sub}})$.
For BCH{[}31,26,3{]}, the difference between LP-UB$(\mathrm{\mathcal{K}_{sub}})$
and LPD$(\mathrm{\mathcal{K}_{sub}})$ is about 0.47 dB while ILP-UB$(\mathrm{\mathcal{K}_{sub}})$
shows an improvement of 0.13 dB. And for BCH{[}63,57,3{]}, the difference
between LP-UB$(\mathrm{\mathcal{K}_{sub}})$ and LPD$(\mathrm{\mathcal{K}_{sub}})$
is about 0.62 dB while ILP-UB$(\mathrm{\mathcal{K}_{sub}})$ shows
an improvement of 0.16 dB. 

The results of the LDPC Tanner code were omitted, since the improvement
of the ILP-UB$(\mathrm{\mathcal{K}_{sub}})$ over the LP-UB$(\mathrm{\mathcal{K}_{sub}})$
at error rate of $10^{\lyxmathsym{\textminus}3}$ is dropped to about
0.05 dB. The reason for that is twofold. First, the Tanner code has
a large average angle: $35.16^{\circ}$. Second, the generators have
an high pseudo-weight: $w_{p}^{min}\approx16.403$. These two values
are high as compared to the other tested codes. 

\begin{flushleft}
\begin{figure}[H]
\begin{raggedleft}
\begin{minipage}[c][1\totalheight][t]{0.45\textwidth}%
\selectlanguage{british}%
\begin{flushleft}
\subfloat[{\selectlanguage{english}%
{\scriptsize \label{fig:ILP-UB  HG''} }\fontsize{9}{10}\selectfont
Golay{[}24,12,8{]} code: results for $231$ generators with $w_{p}\leq3.25$
($w_{p}^{min}=3.2$).\selectlanguage{british}%
}]{\selectlanguage{english}%
\begin{centering}
\includegraphics[scale=0.1]{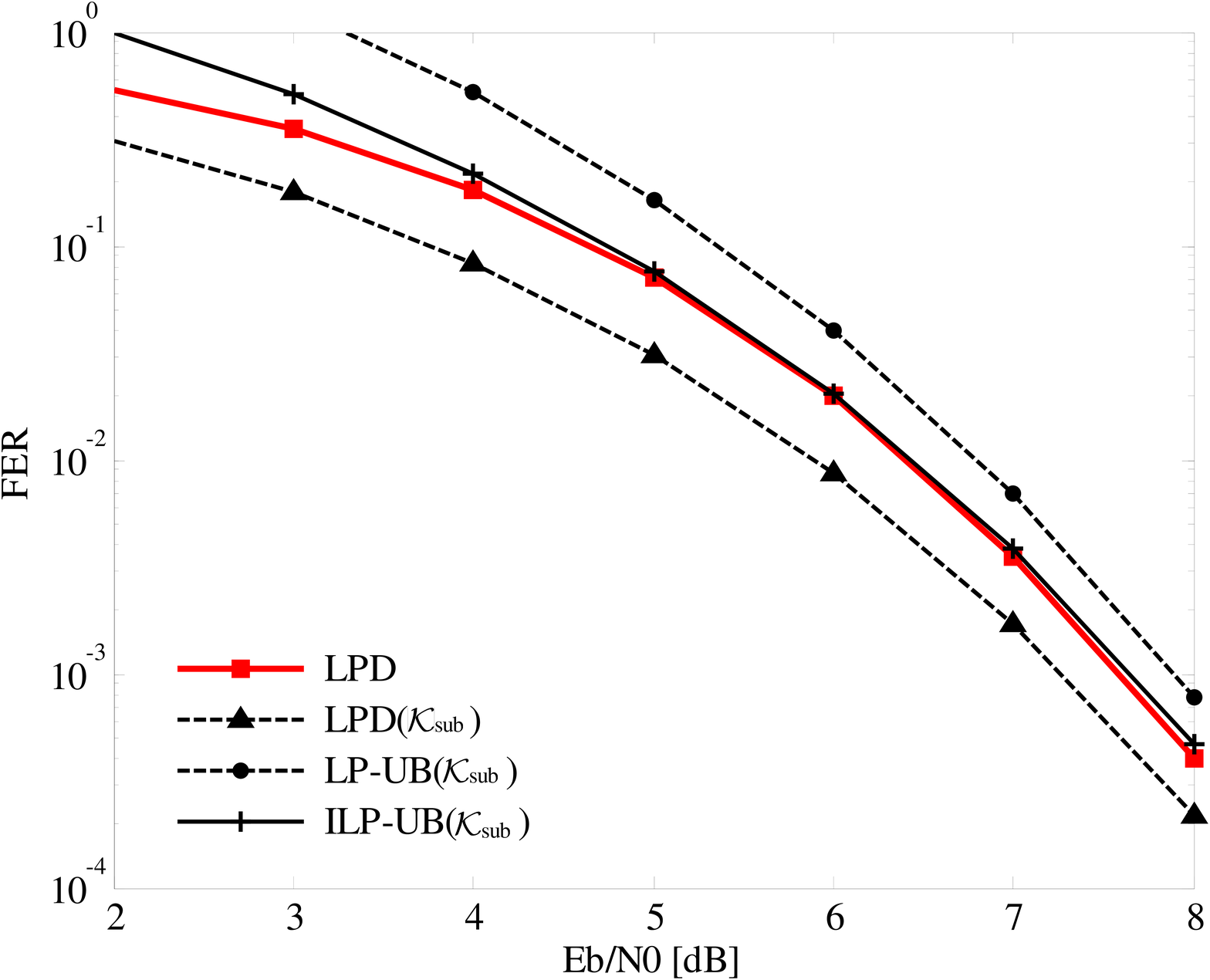}
\par\end{centering}

\selectlanguage{british}%
}
\par\end{flushleft}\selectlanguage{english}%
\end{minipage}\hfill{}%
\begin{minipage}[c][1\totalheight][t]{0.45\textwidth}%
\begin{flushright}
\smallskip{}
\foreignlanguage{british}{}\subfloat[{{\scriptsize \label{fig:BCH(31,26) ILP-UB} }\fontsize{9}{10}\selectfont
BCH{[}31,26,3{]} code: results of all the $1,185$ minimal-weight
generators ($w_{p}^{min}=w_{H}^{min}=3$).}]{\includegraphics[scale=0.1]{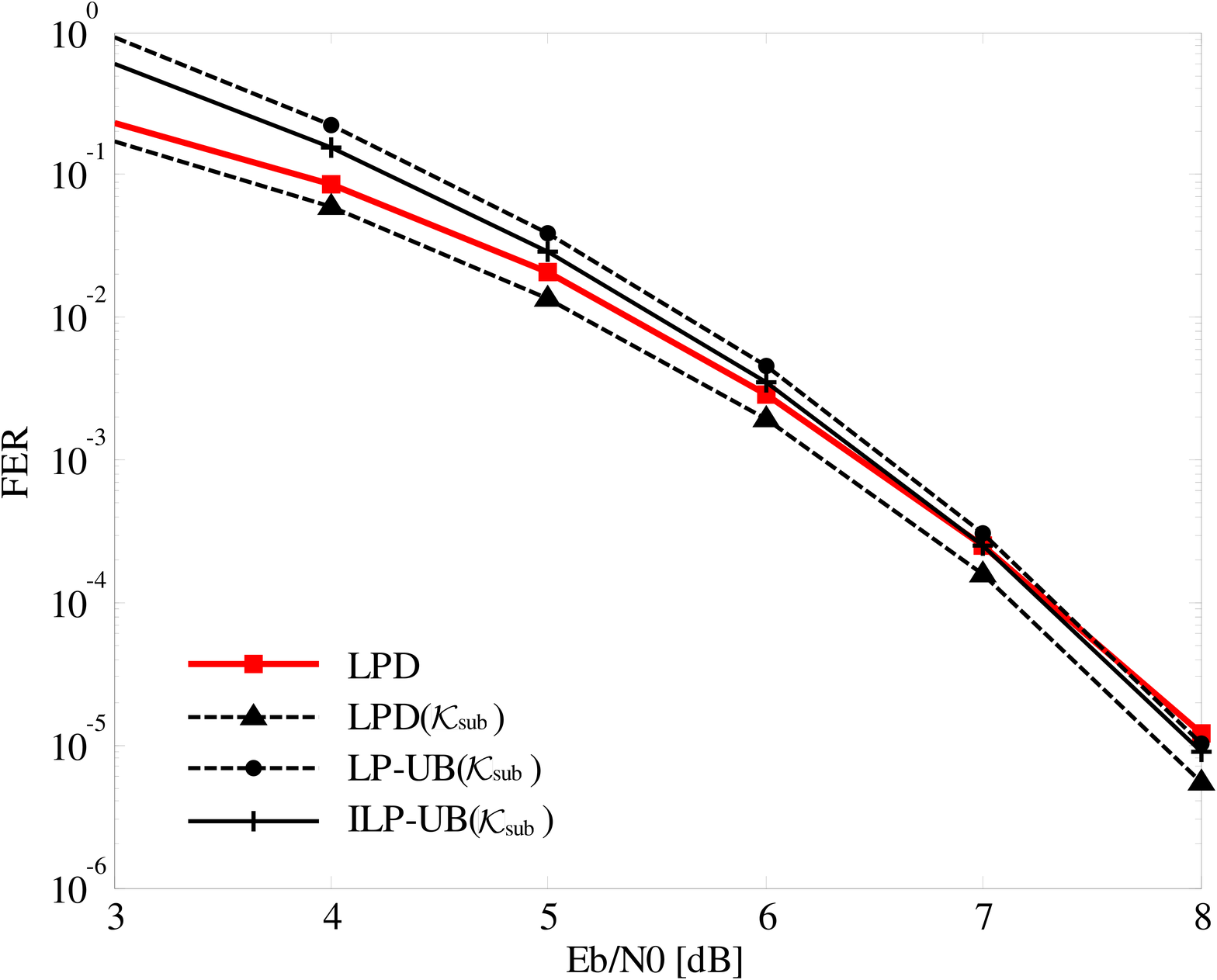}

\selectlanguage{british}%
\raggedright{}\selectlanguage{english}%
}
\par\end{flushright}%
\end{minipage}\hfill{}
\par\end{raggedleft}

\centering{}%
\begin{minipage}[c][1\totalheight][t]{0.45\textwidth}%
\selectlanguage{british}%
\begin{flushleft}
\subfloat[{\selectlanguage{english}%
{\footnotesize \label{fig:BCH(63,57) ILP-UB}}\fontsize{9}{10}\selectfont
BCH{[}63,57,3{]} code: results of all the $11,551$ minimal-weight
generators ($w_{p}^{min}=w_{H}^{min}=3$).\selectlanguage{british}%
}]{\selectlanguage{english}%
\begin{centering}
\includegraphics[scale=0.1]{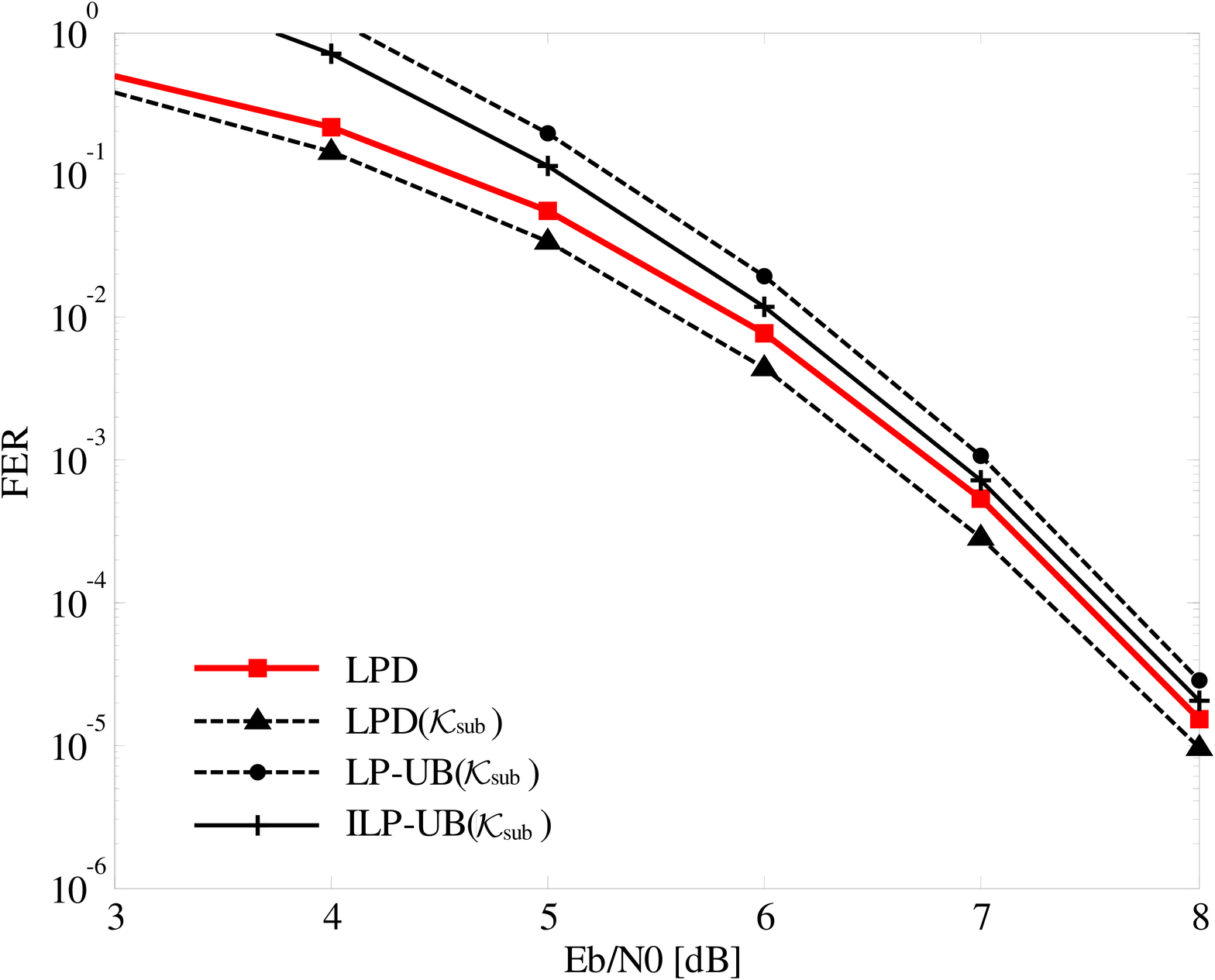}
\par\end{centering}

\selectlanguage{british}%
\raggedleft{}}
\par\end{flushleft}\selectlanguage{english}%
\end{minipage}\caption{\label{fig::ResultesILP-UB}\fontsize{9}{10}\selectfont A comparison
between ILP-UB, LP-UB, LPD and LPD FER for HDPC codes.}
\end{figure}

\par\end{flushleft}

Fig . \ref{fig::ResultesILP-UB} together with Fig. \ref{fig:ResultesAngledistribution}
show that the lower the average angle is, the more improvement the
ILP-UB has. A small average angle is typical for HDPC codes, therefore,
the advantage of ILP-UB over the LP-UB will be reflected better on
such type of codes. But on the other hand, as the larger the average
angle is, the better the LP-UB will be. Fig. \ref{fig:ILP-UB  HG''}
presents the highest improvement of the ILP-UB$(\mathrm{\mathcal{K}_{sub}})$
among the other codes. This result correlates to Golay's smallest
average angle: $19.85^{\circ}$. However, it presents the largest
gap to its $\mathrm{LPD}(\mathrm{\mathcal{K}_{sub}})$. This apparently
happens because there are a significant probabilities of intersections
between three error events or more.

Buksz\'ar and Pr\'ekopa have suggested \cite{CherryTrees} a third
order upper bound on the probability of a finite union of events.
Their bound considers intersections of two and three events. They
proved that this third order bound, which is obtained by the use of
a type of graph called cherry tree, is at least as strong as the second-order
bound. Therefore, implementing such a bound will improve (or at least
will be equal to) the proposed ILP-UB.

\section{Conclusions\label{Conclusions} }

In this paper, we have presented an improved union bound on the error
probability of LP decoding of binary linear HDPC codes transmitted
over a binary-input AWGN channel. It is based on the second-order
upper bound on the probability of a finite union of events. It has
low computational complexity since it only involves the Q-function.
It can be implemented with running time of $O(|\mathcal{G}|^{2})$,
where $\mathcal{G}$ is a set of generators of the fundamental cone
arisen from a given parity check matrix. We examined the proposed
bound for several HDPC codes: Golay{[}24,12,8{]}, BCH{[}31,26,3{]},
BCH{[}63,57,3{]}, and for the LDPC Tanner {[}155,64,20{]} code. The
improvement of the proposed bound over the union bound presents dependency
on the pseudo-weight of the generators and their density. We studied
and compared the generator density through the angle distribution
of various codes and parity-check matrices. Finally, a third order
upper bound was proposed, it is based on a type of graph called cherry
tree, and is left open for further research.



\end{document}